\documentclass[usenatbib]{mnras}

\usepackage{float}

\usepackage[T1]{fontenc}
\usepackage{comment}
\usepackage{scalerel}

\usepackage{mathtools,hyperref}
\usepackage{amsmath}
\usepackage{graphicx}
\usepackage[figuresright]{rotating}
\usepackage{natbib}
\usepackage[all]{hypcap}								
\usepackage{fancyhdr}
\usepackage{bm}
\usepackage{url}
\usepackage{epsfig}
\usepackage{amsmath}
\usepackage{verbatim}

\DeclareRobustCommand{\VAN}[3]{#2}
\let\VANthebibliography\thebibliography
\def\thebibliography{\DeclareRobustCommand{\VAN}[3]{##3}\VANthebibliography}

\usepackage{graphicx}	
\usepackage{amsmath}	
\usepackage{amssymb}	
\usepackage{dsfont}

\usepackage{newtxtext,newtxmath}
\usepackage[dvipsnames]{xcolor}

\newcommand{\mpagano}[1]{{\color{Periwinkle} \textbf{[MP:  #1]}}}

\title[Inpaint Residuals Characterization]{Characterization Of Inpaint Residuals In Interferometric Measurements of the Epoch Of Reionization}

\author[M.~Pagano et al.]{	Michael Pagano$^{1}$\thanks{Email: michael.pagano@mcgill.ca},
	Jing Liu$^{1}$,
	Adrian  Liu$^{1}$,
	Nicholas S. Kern$^{3,23}$,
	Aaron  Ewall-Wice$^{3,16}$,
	\newauthor
	Philip  Bull$^{11,12}$,
	Robert Pascua$^{1}$,
	Siamak Ravanbakhsh$^1$,
	Zara  Abdurashidova$^{3}$,
\newauthor
	Tyrone  Adams$^{4}$,
	James E. Aguirre$^{5}$,
	Paul  Alexander$^{6}$,
	Zaki S. Ali$^{3}$,
	Rushelle  Baartman$^{4}$,
\newauthor
	Yanga  Balfour$^{4}$,
	Adam P. Beardsley$^{2,7}$,
	Gianni  Bernardi$^{8,9,4}$,
	Tashalee S. Billings$^{5}$,
\newauthor
	Judd D. Bowman$^{2}$,
	Richard F. Bradley$^{10}$,
	Jacob  Burba$^{13}$,
	Steven  Carey$^{6}$,
\newauthor
	Chris L. Carilli$^{14}$,
	Carina  Cheng$^{3}$,
	David R. DeBoer$^{15}$,
	Eloy  de~Lera~Acedo$^{6}$,
	Matt  Dexter$^{15}$,
\newauthor
	Joshua S. Dillon$^{3}$,
	Nico  Eksteen$^{4}$,
	John  Ely$^{6}$,
	Nicolas  Fagnoni$^{6}$,
\newauthor
	Randall  Fritz$^{4}$,
	Steven R. Furlanetto$^{17}$,
	Kingsley  Gale-Sides$^{6}$,
	Brian  Glendenning$^{18}$,
\newauthor
	Deepthi  Gorthi$^{3}$,
	Bradley  Greig$^{19}$,
	Jasper  Grobbelaar$^{4}$,
	Ziyaad  Halday$^{4}$,
\newauthor
	Bryna J. Hazelton$^{20,21}$,
	Jacqueline N. Hewitt$^{22,23}$,
	Jack  Hickish$^{15}$,
	Daniel C. Jacobs$^{2}$,
\newauthor
	Austin  Julius$^{4}$,
	MacCalvin  Kariseb$^{4}$,
	Joshua  Kerrigan$^{13}$,
\newauthor
	Piyanat  Kittiwisit$^{12}$,
	Saul A. Kohn$^{5}$,
	Matthew  Kolopanis$^{2}$,
	Adam  Lanman$^{13}$,
	Paul  La~Plante$^{3,5}$,
\newauthor
	Anita  Loots$^{4}$,
	David Harold~Edward MacMahon$^{15}$,
	Lourence  Malan$^{4}$,
	Cresshim  Malgas$^{4}$,
\newauthor
	Keith  Malgas$^{4}$,
	Bradley  Marero$^{4}$,
	Zachary E. Martinot$^{5}$,
	Andrei  Mesinger$^{24}$,
\newauthor
	Mathakane  Molewa$^{4}$,
	Miguel F. Morales$^{20}$,
	Tshegofalang  Mosiane$^{4}$,
	Abraham R. Neben$^{23}$,
\newauthor
	Bojan  Nikolic$^{6}$,
	Hans  Nuwegeld$^{4}$,
	Aaron R. Parsons$^{3}$,
	Nipanjana  Patra$^{3}$,
	Samantha  Pieterse$^{4}$,
\newauthor
	Nima  Razavi-Ghods$^{6}$,
	James  Robnett$^{14}$,
	Kathryn  Rosie$^{4}$,
	Peter Sims$^{1}$,
\newauthor
	Craig  Smith$^{4}$,
	Hilton  Swarts$^{4}$,
	Nithyanandan  Thyagarajan$^{25,14}$,
	Pieter  van~Wyngaarden$^{4}$,
\newauthor
	Peter K.~G. Williams$^{26,27}$,
	Haoxuan  Zheng$^{23}$
\\
$^{1}$ Department of Physics and McGill Space Institute, McGill University, 3600 University Street, Montreal, QC H3A 2T8, Canada\\
$^{2}$ School of Earth and Space Exploration, Arizona State University, Tempe, AZ\\
$^{3}$ Department of Astronomy, University of California, Berkeley, CA\\
$^{4}$ South African Radio Astronomy Observatory, Black River Park, 2 Fir Street, Observatory, Cape Town, 7925, South Africa\\
$^{5}$ Department of Physics and Astronomy, University of Pennsylvania, Philadelphia, PA\\
$^{6}$ Cavendish Astrophysics, University of Cambridge, Cambridge, UK\\
$^{7}$ Department of Physics, Winona State University, Winona, MN\\
$^{8}$ INAF-Istituto di Radioastronomia, via Gobetti 101, 40129 Bologna, Italy\\
$^{9}$ Department of Physics and Electronics, Rhodes University, PO Box 94, Grahamstown, 6140, South Africa\\
$^{10}$ National Radio Astronomy Observatory, Charlottesville, VA\\
$^{11}$ Jodrell Bank Centre for Astrophysics, University of Manchester, Manchester, M13 9PL, UK\\
$^{12}$ Department of Physics and Astronomy,  University of Western Cape, Cape Town, 7535, South Africa\\
$^{13}$ Department of Physics, Brown University, Providence, RI\\
$^{14}$ National Radio Astronomy Observatory, Socorro, NM 87801, USA\\
$^{15}$ Radio Astronomy Lab, University of California, Berkeley, CA\\
$^{16}$ Department of Physics, University of California, Berkeley, CA\\
$^{17}$ Department of Physics and Astronomy, University of California, Los Angeles, CA\\
$^{18}$ National Radio Astronomy Observatory, Socorro, NM\\
$^{19}$ School of Physics, University of Melbourne, Parkville, VIC 3010, Australia\\
$^{20}$ Department of Physics, University of Washington, Seattle, WA\\
$^{21}$ eScience Institute, University of Washington, Seattle, WA\\
$^{22}$ MIT Kavli Institute, Massachusetts Institute of Technology, Cambridge, MA\\
$^{23}$ Department of Physics, Massachusetts Institute of Technology, Cambridge, MA\\
$^{24}$ Scuola Normale Superiore, 56126 Pisa, PI, Italy\\
$^{25}$ Commonwealth Scientific and Industrial Research Organisation (CSIRO), Space \& Astronomy, P. O. Box 1130, Bentley, WA 6102, Australia\\
$^{26}$ Center for Astrophysics, Harvard \& Smithsonian, Cambridge, MA\\
$^{27}$ American Astronomical Society, Washington, DC\\
}

\begin{document}
\label{firstpage}
\pagerange{\pageref{firstpage}--\pageref{lastpage}}
\maketitle
\begin{abstract}
To mitigate the effects of Radio Frequency Interference (RFI) on the data analysis pipelines of 21cm interferometric instruments, numerous inpaint techniques have been developed. In this paper we examine the qualitative and quantitative errors introduced into the visibilities and power spectrum due to inpainting. We perform our analysis on simulated data as well as real data from the Hydrogen Epoch of Reionization Array (HERA) Phase 1 upper limits. We also introduce a convolutional neural network that is capable of inpainting RFI corrupted data. We train our network on simulated data and show that our network is capable at inpainting real data without requiring to be retrained. We find that techniques that incorporate high wavenumbers in delay space in their modeling are best suited for inpainting over narrowband RFI. We show that with our fiducial parameters Discrete Prolate Spheroidal Sequences (DPSS) and CLEAN provide the best performance for intermittent RFI while Gaussian Progress Regression (GPR) and Least Squares Spectral Analysis (LSSA) provide the best performance for larger RFI gaps. However we caution that these qualitative conclusions are sensitive to the chosen hyperparameters of each inpainting technique.  We show that all inpainting techniques reliably reproduce foreground dominated modes in the power spectrum. Since the inpainting techniques should not be capable of reproducing noise realizations, we find that the largest errors occur in the noise dominated delay modes. We show that as the noise level of the data comes down, CLEAN and DPSS are most capable of reproducing the fine frequency structure in the visibilities.
 
\end{abstract}




\section{Introduction}

The Epoch of Reionization (EoR) plays a crucial role in the evolution of the Universe since it is the period in which the intergalactic medium (IGM) transitions from neutral to ionized. The precise details of how the EoR unfold are currently observationally unconstrained. In most models of the EoR, the onset of the first generation galaxies give rise to ionizing photons which gradually disperse across the IGM and ionize the neutral hydrogen marking the beginning of the EoR \citep{ReviewPritchardLoeb, ReviewAdrian2020, ReviewFurlanetto2006, ReviewMorales2010}. One method to directly measure the neutral hydrogen in the IGM during the EoR is to use the 21cm hyperfine transition of hydrogen in which a 21cm wavelength photon is released when the electron flips its spin relative to the proton \citep{Mandau1997, Furlanetto2004, Furlanetto2008}. Thus the 21cm line directly probes the neutral hydrogen in the IGM during the EoR. The emitted 21cm-wavelength photon is then redshifted into radio wavelengths and is potentially observable in contrast to the CMB, enabling tomographic measurements of neutral hydrogen. Ground based interferometric instruments such as the Hydrogen Epoch of Reionization array (HERA) \citep{HERA}, Square Kilometer Array (SKA) \citep{SKAintro}, Precision Array for Probing the Epoch of Re-ionization (PAPER) \citep{PAPERintro} , Murchison Widefield Array (MWA) \citep{MWAintro}, Low Frequency Array (LOFAR) \citep{LOFARintro} have the ability to measure the spatial fluctuations of the 21cm line.

One of the challenges in measuring radio photons using ground based instruments is the frequent data flagging due to radio frequency interference (RFI). Most RFI sources are due to terrestrial transmitters and satellites which lead to narrowband flagging in the data analysis. Other wideband sources of RFI, such as communication satellites, require flagging more substantive portions of the raw data. The excision of RFI in the data analysis introduces gaps in the data which cause artifacts in the 21cm power spectrum. Data analysis pipelines which try to separate the foregrounds from the cosmological signal in the Fourier domain will be directly affected by the RFI gaps in the data. This impedes measurement of the EoR (for example see \cite{Wilensky}). A conservative approach to mitigate the effect of RFI on the power spectrum is to avoid all frequency bands where RFI has corrupted data which ensures that there aren't artifacts in the power spectrum. Doing so severely restricts the available frequency channels to use as part of our analysis thereby preventing us from accessing all redshifts. Further, this approach is not ideal since it decreases the signal to noise of the measurement.

 Data analysis pipelines which are affected by RFI use ``inpainting'' techniques to partially restore the RFI corrupted data. A number of algorithms have been developed to perform inpainting, most notably the CLEAN algorithm which was originally introduced in \cite{CLEANHog}. Although bearing the same name, we use a modified version of CLEAN to fit the inpainting needs in the HERA data analysis pipeline \citep{Parsons2009CLEANvisib}. Besides CLEAN, other inpainting techniques have been explored as well such as least square spectral analysis (LSSA), Gaussian process regression (GPR) \citep{Ghosh2020, Kern2021} and discrete prolate spheroidal sequence (DPSS) \citep{DPSSintro, EW21}. These inpainting methods use the uncorrupted data to form a crude model for the corrupted data which is then replaced into the RFI flagged regions, thereby reducing the effect that RFI has on the 21cm power spectrum. However, the crudely restored data are imperfect and thus they too introduce errors in the analysis. In this paper we critically evaluate the performance of existing inpainting techniques CLEAN, LSSA, GPR, and DPSS in reconstructing corrupted visibility data.  In this paper we study the HERA implementations of these inpainting techniques however similar variations of these techniques have been implemented in other instruments such as \cite{Offringa2019} in the LOFAR experiment and \cite{BarryMWAls} in the MWA. Outside of 21cm cosmology, inpainting has been frequently done in CMB studies \citep{Strack2013, GruetjenCMB, TrottJordanGPR} and gravitational waves analyses \citep{ZackayGWInpainting}.
 
 We also introduce a Convolutional Neural Network (CNN) dubbed as ``U-Paint'' as an alternative to inpainting RFI corrupted data. CNNs have been previously explored as an inpainting technique by \cite{MLInpaintNotUnet, MLInpaintNotUnet2, MLInpaintUnet,MLInpaintUnet2,MLInpaintUnet3,MLInpaintUnet4} but not in the context of radio astronomy experiments. U-Paint marks the introduction of CNNs as an inpainting technique in the data analysis pipelines of radio astronomy. By assessing its effectiveness as compared to existing techniques we show that convolutional neural networks show great promise as an inpainting technique. Using a series of Monte Carlo realizations, we propagate the errors of the inpainted visibilities through to the 21cm power spectrum. We quantify the performance of each inpainting technique and parametrize their errors in the power spectrum. We perform our analysis using the HERA instrument; however, our approach is general enough to apply to any interferometer. This paper is structured as follows. In Section \ref{sec:HERAObservations} we introduce our fiducial instrument HERA as well as sources of RFI which affect the data analysis pipeline. In Section \ref{sec:InpaintingTechniques} we discuss existing inpainting techniques CLEAN, LSSA, GPR, and DPSS as well as quantifying their performance in inpainting corrupted visibilities. In Section \ref{sssec:Unets} we introduce U-Paint which we use to inpaint corrupted data. In Section \ref{sec:BigSectionErrorQuantification} we assess its performance relative to existing inpainting methods. In Section \ref{sec:PowerSpectrumErrorCharacterization} we propagate the inpainting errors through the analysis and characterize their effect on the power spectrum. In Section \ref{sec:RealDataBigSection} we apply our analysis on real HERA data. We conclude in Section \ref{sec:Conclusion}.

    

\section{HERA Observations}
\label{sec:HERAObservations}
In this section we introduce the HERA instrument, an interferometer located in the Karoo desert designed to measure the 21cm power spectrum during Cosmic Dawn and the EoR. Though we use the HERA instrument as the test-bed for analysis, our results and procedures are not strictly limited to HERA and are thus applicable to any interferometer. When completed, HERA will be comprised of 350 14m dishes capable of observing at frequencies 50MHz to 225MHz. In this paper however, we consider the instrumental parameters taken from Phase 1 data used to set the recent HERA upper limits \cite{HERAH1CData} which span frequencies $100$MHz to $200$MHz in 1024 channels using 39 dishes. In this Section we review the data analysis pipeline established in HERA's Phase 1 upper limits, which we use in this paper for consistency. In doing so, we establish notation for the remainder of this paper. We begin in Section \ref{sssec:PowerSpectrum} where we discuss the Phase 1 data analysis pipeline from \cite{HERAH1CData} while in Section \ref{sssec:RFI_introduction} we discuss RFI scenarios which affect interferometric measurements at low frequencies. In Section \ref{sssec:P1V_Data} we discuss the simulated datasets that we use as part of our analysis as well as real data from the Phase 1 data release.
\subsection{RFI Flagging}
\label{sssec:RFI_introduction}
\begin{figure}
  \includegraphics[width=0.5\textwidth]{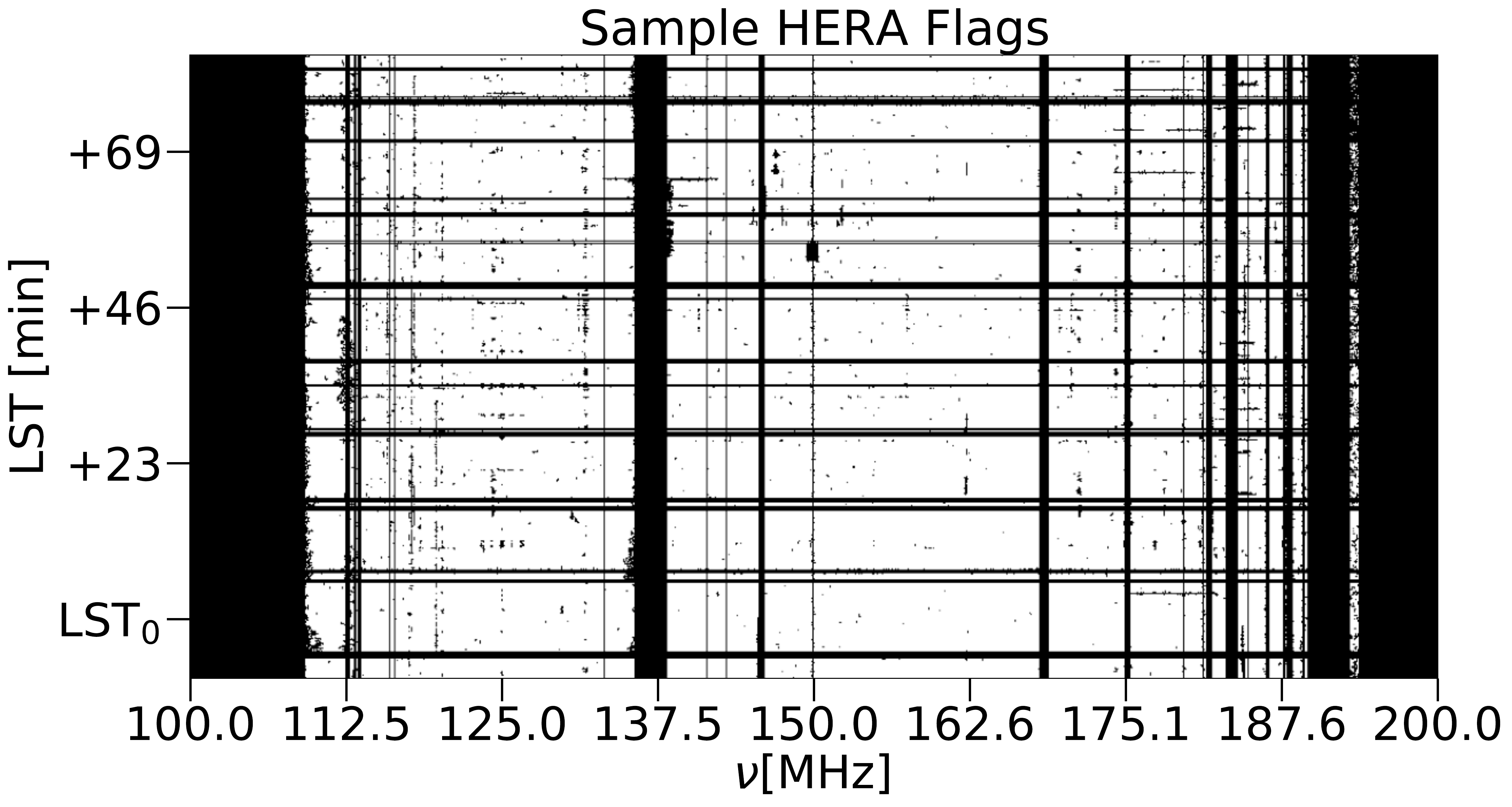}
  \caption{Sample HERA flags split from 100MHz-200MHz. Frequency channels below $110$MHz are reserved for FM radio. The ORBCOMM satellite is responsible for RFI at $\nu = 136$MHz. Frequency channels above $\nu = 174$MHz are flagged due to broadcast television. }
  \label{fig:HERAFlags}
\end{figure}

Though we discuss the effect of RFI on our fiducial instrument HERA, the systematics caused by RFI are equally applicable to other instruments. Radio experiments located on the ground ubiquitously experience RFI. The origin of the RFI are either terrestrial in nature or due to satellites. Terrestrial sources can range from cell-phones, WiFi as well any other radio producing mechanism sourced on the ground. This includes FM radio and broadcast television. The amount of terrestrial RFI can be minimized by operating the instrument in radio quiet zone, such as the Karoo desert in HERA's case. This minimizes terrestrial RFI but does not totally eliminate it \citep{KerriganMLOptimization, HERAH1CData, HERAflaggingKohn2016, PaulLPHERArfi, ChenPaulDeepRFI} .
For brevity, we find it useful to organize RFI by the number of frequency channels they occupy. We shall denote RFI which occupies relatively few channels ($\sim 1-3$) as narrowband RFI. We assign the RFI to be wideband if it occupies a more significant fraction of the frequency band. Note that we are not setting a strict definition of narrowband or wideband RFI, rather we find it convenient to use this notation in our analysis. In Figure \ref{fig:HERAFlags} we show example HERA flags. The most frequent type of RFI are narrowband emitters which can occur irregularly in $\nu$ and $t$ creating a scattered assortment of flags in the visibilities. However other wideband types of RFI can occur more predictably in the dataset. For example,  ORBCOMM satellite communication at $\nu = 136-138$MHz, broadcast television at $\nu > 174$MHz. While a FM radio broadcast occupies a single frequency channel, frequencies $\nu < 111$MHz are reserved for FM broadcast. 

HERA searches for RFI in the visibilities by scanning the data for localized irregularities. Adjacent data in $\nu$ and local sidereal time (LST) are used to differentiate between RFI and thermal noise fluctuations. This procedure is applied after the absolute calibration step of the visibilities so that any issues with the instrument can also be flagged (see Figure 3 in  \cite{HERAH1CData} for a detailed description of the HERA data analysis pipeline). For example, in this flagging scheme, intermittent correlator integration failures (a source of wideband flags) can also be flagged. The LST binned visibilites are also manually scanned for narrowband RFI that was undetected by the automated flagging process. 

\subsection{Power Spectrum}
\label{sssec:PowerSpectrum}
HERA Phase 1 observed the radio sky at frequencies $100$MHz to $200$MHz over 1024 channels corresponding to a channel width of $\Delta \nu \simeq 0.1$MHz. These frequencies are measured at time cadence of $\Delta t = 10.7$s.  The raw data taken from correlated antennas in the interferometer are termed the visibilities $V$, which depend on the observation frequency $\nu$, and the time of observation ``LST''. The visibilities are complex valued and thus can be expressed either in terms of their real and imaginary components or amplitude and phase. We denote the amplitude of the visibilities as $|V|$ and the phase of the visibilities as $\phi$. Since the visibilities are the product of correlated antennas, the visibilities are simultaneously measured on all antenna combinations within the HERA antenna array. The visibilities measured by the HERA interferometer using the $i$th antenna at position ${\bf x}_i$ and $j$th antenna at position ${\bf x}_j$ form a baseline ${\bf b} = {\bf x}_i - {\bf x}_j$. It was shown by \cite{Parsons2009CLEANvisib, ParsonsPoberAguirre2012} that for a single baseline $\bf{b}$ at observation frequency $\nu$, the visibilities can be written as
\begin{equation}
    \label{eq:visibilityDefinition}
    V(u,v) = \int dl dm A(l,m,\nu)T(l,m,\nu,t)e^{-2\pi i \nu \tau_g} 
\end{equation}
where $A(l,m)$ is the primary beam of the instrument and $T(l,m)$ is the temperature of the sky. The time dependence arises because the sky rotates above the instrument. The terms $l\equiv sin(\theta_x)$ and $m\equiv sin(\theta_y)$ encode the angular components of the sky and $\tau_g$ is given by
\begin{equation}
    \label{eq:tau_g1}
    \tau_g \equiv \frac{\bf{b}\cdot \hat{s}}{c} = \frac{1}{c}\left ( b_xl +b_ym +b_z\sqrt{1- l^2 - m^2} \right )
\end{equation}
where $\tau_g$ is the geometric delay corresponding to the projection of the baseline ${\bf b} = (b_x, b_y, b_z)$ in the direction $\hat{s} = (l,m,\sqrt{1- l^2 - m^2})$ and where $c$ is the speed of light. Although the baseline $\bf{b}$ in Equation \ref{eq:tau_g1} can represent any antenna pairing in the HERA array, in this paper we focus our analysis to only the shortest baselines, i.e. adjacent antenna pairs. The Fourier transform of the visibilities in Equation \ref{eq:visibilityDefinition} along the frequency direction is defined as 
\begin{equation}
    \label{eq:Vtilde}
    \widetilde{V}(\tau, t) = \int d\nu dl dm A(l,m,\nu)  T(l,m,\nu,t) \phi(\nu) e^{2\pi i \nu (\tau - \tau_g)}
\end{equation}

where $\tau$ is the Fourier dual to frequency in the Fourier transform called the delay. The term $\phi(\nu)$ denotes a tapering function that defines our spectral window of observation. For consistency with analysis from the Phase 1 upper limits, we use the Blackman-Harris window function as our tapering function $\phi(\nu)$. The delay power spectrum can be estimated by the square of $\widetilde{V}(\textbf{b}, \tau)$:

\begin{equation}
    \label{eq:delay_pspec}
    P( k_{\perp}, k_{\parallel} )  = \frac{X^2Y}{\Omega_{\rm{pp}}B} \left | \widetilde{V}(\textbf{u} , \tau) \right |^2
\end{equation}
where $k_{\perp}$ is the wavenumber corresponding to the plane of the sky and $k_{\parallel}$ parallel to the line of sight. The visibility coordinates $\boldsymbol u $ are related to the frequency $\nu$ through $\boldsymbol u = \nu \bf{b}/c$. The term $\Omega_{\rm{pp}}$ gives the angular area by integrating the square of the primary beam, while $B$ is an effective bandwidth given by $\int d\nu |\phi|^2$. The term $k_{\perp}$ can be related to the baseline $\textbf{b}$ using $k_{\perp} = \frac{2\pi \nu \textbf{b}}{cX}$. The term $k_\parallel$ can be written as $k_{\parallel} = \frac{2\pi \tau }{Y}$ where $\tau$ is the Fourier dual to the frequency axis $\nu$ with dimensions of $1/\nu$. The factor $X$ converts comoving distance $r_{\perp}$ to angular separation $\theta$, while $Y$ converts radial comoving distances $r_{\parallel}$ to frequency intervals $\Delta \nu$:
\begin{align}
    \label{eq:XandY}
    X  \equiv& \frac{r_\perp}{\theta} = \frac{c}{H_0}\int^z_0 \frac{dz'}{E(z')}\\
    Y \equiv& \frac{\Delta r_\parallel}{\Delta \nu} = \frac{c}{H_0 \nu_{21}} \frac{(1+z)^2}{E(z)}
\end{align}
and where $H_0$ is the Hubble parameter, $E(z) \equiv \sqrt{\Omega_m(1+z)^3 + \Omega_\Lambda}$ and $\Omega_\Lambda$ the normalized dark energy density and $\nu_{\rm 21} \approx 1420$MHz, the rest frequency of the 21cm line. 
For a drift scan telescope like HERA, one typically first averages $ \widetilde{V}(\textbf{u} , \tau) $ at identical LSTs across different sidereal days.
This process is referred to as coherent averaging. Once the power spectrum of the coherently averaged visibilities is computed, one then averages $P( k_\perp, k_{\parallel} )$ across different LSTs, a process known as incoherent averaging. In an observationally realistic data analysis pipeline (i.e. that aims to measure cosmological signal), instead of directly computing $P( k_\perp, k_{\parallel} )$ using Equation \ref{eq:delay_pspec}, one instead forms the cross spectra using different times or baselines in order to avoid a noise bias. In this scenario one forms the product of the visibilities at different times or baselines within the context of Equation \ref{eq:delay_pspec}. Since the objective of this paper is to characterize the statistical properties of inpaint models, and not to measure cosmological signal, we do not form the cross-spectra as described above. Thus, the noise bias will be present in our estimates of power spectra. To evaluate the power spectrum in Equation \ref{eq:delay_pspec}, we use the publicly available code \texttt{hera} \texttt{pspec}\footnote{https://github.com/HERA-Team/hera\_pspec}.

The delay power spectrum in Equation \ref{eq:delay_pspec} is dominated by galactic and  extra-galactic sources of radio emission referred to as the ``foregrounds''. The foregrounds are orders of magnitude brighter than the anticipated 21cm signal. The foregrounds are spectrally smooth, and thus can be crudely approximated by a flat spectum. Under this assumption the temperature of the sky in Equation \ref{eq:visibilityDefinition} loses its dependence on frequency, $T(l,m,\nu,t) \simeq T(l,m,t)$. If the beam and spectral window are also frequency independent, with a infinitely large bandpass then the delay $\tau$ in Equation \ref{eq:delay_pspec} is geometrically limited by the baseline length $\bf{b}$ and the speed of light to values:
\begin{equation}
    \label{eq:Maximum_delay}
    \tau_{\rm g} \le \frac{| \boldsymbol b |}{c} .
\end{equation}
Under these idealistic assumptions the foregrounds are confined to within $\tau_{g}$; however, since the foregrounds are only approximately smooth as a function of frequency and both the primary beam and $\phi$ are also not frequency independent. Thus the foregrounds spread outside the confines of $\tau_{\rm g}$ \citep{QuantifyingEoRDelaySpectrumAdamLanman}. Though in this paper we separate our analysis for $\tau$ modes inside and outside of $\tau_g$, it should be noted that our analysis is not stringent on the true value of $\tau_g$, rather $\tau_g$ serves as a convenient marker for modes which are mostly dominated by the foregrounds and modes which are relatively foreground free. Also note that in computing the power spectrum (Equation \ref{eq:delay_pspec}) we apply the Blackman-Harris tapering function. Since this operation is a convolution, this spreads power from each bin to neighbouring bins. Thus $\tau_{\rm g}$ modes which are dominated by the foregrounds are spread into adjacent bins. The objective of this work is to establish the errors in the data analysis pipeline due to inpainting. The errors do not strictly depend on which $\tau$ modes are part of the wedge. Thus we conservatively include $\tau$ modes satisfying $|\tau| < 500$ns to capture the spillover of foreground power into neighbouring $\tau$ bins and for brevity we refer to all of these modes as the ``wedge''.  

The presence of flagged channels in the dataset complicates the above power spectrum analysis. Equation \ref{eq:delay_pspec} is a Fourier transform of the visibilities along the frequency direction. Performing a Fourier transform of a dataset which contains masked regions will cause artifacts in the resulting Fourier spectrum. This effect is similar to carrying out a Fourier analysis of a top-hat function which creates a ``ringing'' at high delay modes. We thus expect excess power in the large $\tau$ domain. 
Thus analyses which sample the visibilites in the EoR window at high delay will be especially affected the by the artifacts due to flags in the data.  One conservative approach to circumvent this issue is to avoid frequency channels which have been flagged and select cleaner windows in the visibilites which are unaffected by RFI. This strategy reduces the amount of data in the analysis and thus decreases the signal to noise.  

\subsection{Datasets}
\label{sssec:P1V_Data}
In this Section we introduce the datasets (i.e. visibilities) which we use as part of our analysis. We consider two separate sets of visibilities, real HERA data and simulations of HERA observations. For the simulated visibilities we also consider different noise scenarios.

For the real data we use HERA's phase 1 visibilities (hereafter denoted as P1V) in \cite{HERAH1CData}, we use data from the IDR2 dataset which spans a range of right ascensions from 0 to 12 hours. The instrument parameters match those from Section \ref{sssec:P1V_Data}. Since raw HERA data is propagated through a data analysis pipeline there are a number of places along the pipeline where we might choose to apply our analysis. We choose to use the visibilities after they have been absolutely calibrated. Our primary motivation for this is because the LST binning process results in averaging the visibilities by the number of observation nights resulting in lower noise. This makes it slightly easier for the inpainting algorithms due to the lower noise and also since there is intermittent RFI that isn't present every day. In future work we can take advantage of the symmetries between visibility data on different days by implementing network changes such as in \cite{DSSintro} which are optimized to take advantage of symmetries in datasets. 

For simulated data, we use the simulations from the HERA validation pipeline in \cite{H1CValidationPaper}. The simulated visibilities in \cite{H1CValidationPaper} are designed to be a realistic representation of the sky as seen through the HERA instrument and thus the instrumental parameters match those of the true visibilities. We briefly review the simulated data here though the reader is encouraged to see \cite{H1CValidationPaper} for further details. To create a model of the sky composed of a foreground, and EoR component are put through a mock HERA observation simulater, \texttt{RIMEz}, an internally developed software which correctly simulates HERA's drift scan capabilities, and is capable of sampling the sky at the cadence of HERA time sampling over HERA's full frequency resolution and bandwidth. Though RIMEz simulation also take into account instrumental effects such as cross-coupling and reflection systematics, we do not include them in our simulations. The sky model is generated by adding an EoR component to the foregrounds. The EoR component is modeled as a Gaussian random temperature field with power spectrum $P_{\rm EoR} = A_0k^{-2}$
where this relationship approximates those which are obtained by simulations and where $A_0$ is the amplitude of the power spectrum. The EoR component is added to foreground model which is composed of GLEAM sources and diffuse emission. Only GLEAM sources with an associated spectral model are considered. The GLEAM sources are composed of approximately $2.4 \times 10^5$ sources in the catalog which\cite{GLEAMsources2017}, each with a power law emission spectrum given by 
\begin{equation}
    \label{eq:GLEAMpLaw}
    I_p(\nu, \boldsymbol{\hat{s}} ) = \sum_{n}^{240\times 10^3} F_n \left (\frac{\nu}{\nu_0} \right )^\beta \delta(1 -  \boldsymbol{\hat{s}} \cdot \boldsymbol{\hat{s}}_n) 
\end{equation}
where $F_n$ is the flux of the $n$th point source, $\beta$ the spectral index which characterizes the power law and $\boldsymbol{\hat{s}}$ is its position. Note that since the GLEAM catalog has coverage gaps in regions within HERA's spatial observation window, the observing times of the simulations are chosen as to avoid times where these gaps coincide with HERA's primary beam. The diffuse emission component of the foregrounds is simulated based on the Global Sky Model in \cite{GSMImprovedMax} and \cite{GSM2008}. Thermal noise is generated and added to the simulations by drawing samples from a Gaussian distribution with zero mean and standard deviation $\hat{\sigma}_0$ that depends on the time and frequency of observation as well as the amplitude of the auto-correlation of each baseline through the radiometer equation
\begin{equation}
    \label{eq:Radiometer}
    \hat{\sigma}_0(\nu, t) = \alpha \frac{ \kappa(\nu)\Omega(\nu) \left ( T_{\rm auto} (\nu, t) + T_{\rm rx}\right )  }{\sqrt{\Delta \nu \Delta t}}
\end{equation}
where $\Delta t$ is the time integration of 10.7s for HERA, $\Delta \nu$  is HERA's channel width, i.e. $\Delta \nu \simeq 0.1$MHz and $T_{\rm rx}$ is the receiver temperature (assumed to be uniform in $\nu$ and independent of antenna, see \cite{H1CValidationPaper} for precise values) in units of K/str. The term $\kappa(\nu)\Omega(\nu)$ is a conversion factor from K/str to Jy through $\kappa(\nu) = (2k_B \times 10^{26})/(A(\nu)\Omega(\nu))$ where $k_B$ is the Boltzmann constant, and $A(\nu)$ is the effective area and $\Omega(\nu)$ is the solid angle of the beam. The parameter $\alpha$ is a dimensionless parameter which we use to simulate scenarios with higher levels of thermal noise. We consider values of $\alpha = \left[1, 2, 3, 4, 7\right]$. In our fiducial noise level $\alpha = 1$. The total simulated visibilities spans roughly 13 hour observations corresponding to over $\gtrsim 4000$ time integrations of 10.7 seconds each. The simulation data is composed of 39 operational antennas with north and east pointing polarisations. We consider only the shortest baselines (i.e. antennas separated by 14.7m) in this work. We find that our results do not depend on the specific antennas used to form the 14.7m baseline. Thus without loss of generality we perform our analysis using the antenna pair $(84,85)$, including multiple linear polarisations ( EE and NN ). We have repeated our subsequent analyses for redundant baselines using other antenna pairs and have found no significant differences in our qualitative or quantitative results. Since this is a simulated dataset, there are not any RFI corrupted regions. To imitate a scenario where RFI has corrupted regions of our simulated visibilities, we apply the HERA flags discussed in the previous section to our dataset.


\section{Inpainting Techniques}
\label{sec:InpaintingTechniques}

In this section we describe the inpainting methods that we use as part of our analysis.  We begin by introducing CLEAN and LSSA in Sections \ref{sssec:CLEAN} and \ref{sssec:LSSA}. In these sections we also compute the optimal value of CLEAN and LSSA hyperparameters to optimize their respective performances. In Section \ref{ssec:inv_cov} we introduce the covariance-Based Inpainting methods, GPR \& DPSS. Finally in Section \ref{sssec:Unets} we introduce the neural network architecture of U-Paint.

\subsection{CLEAN}
\label{sssec:CLEAN}
 \begin{figure*}
  \includegraphics[width=0.75\textwidth]{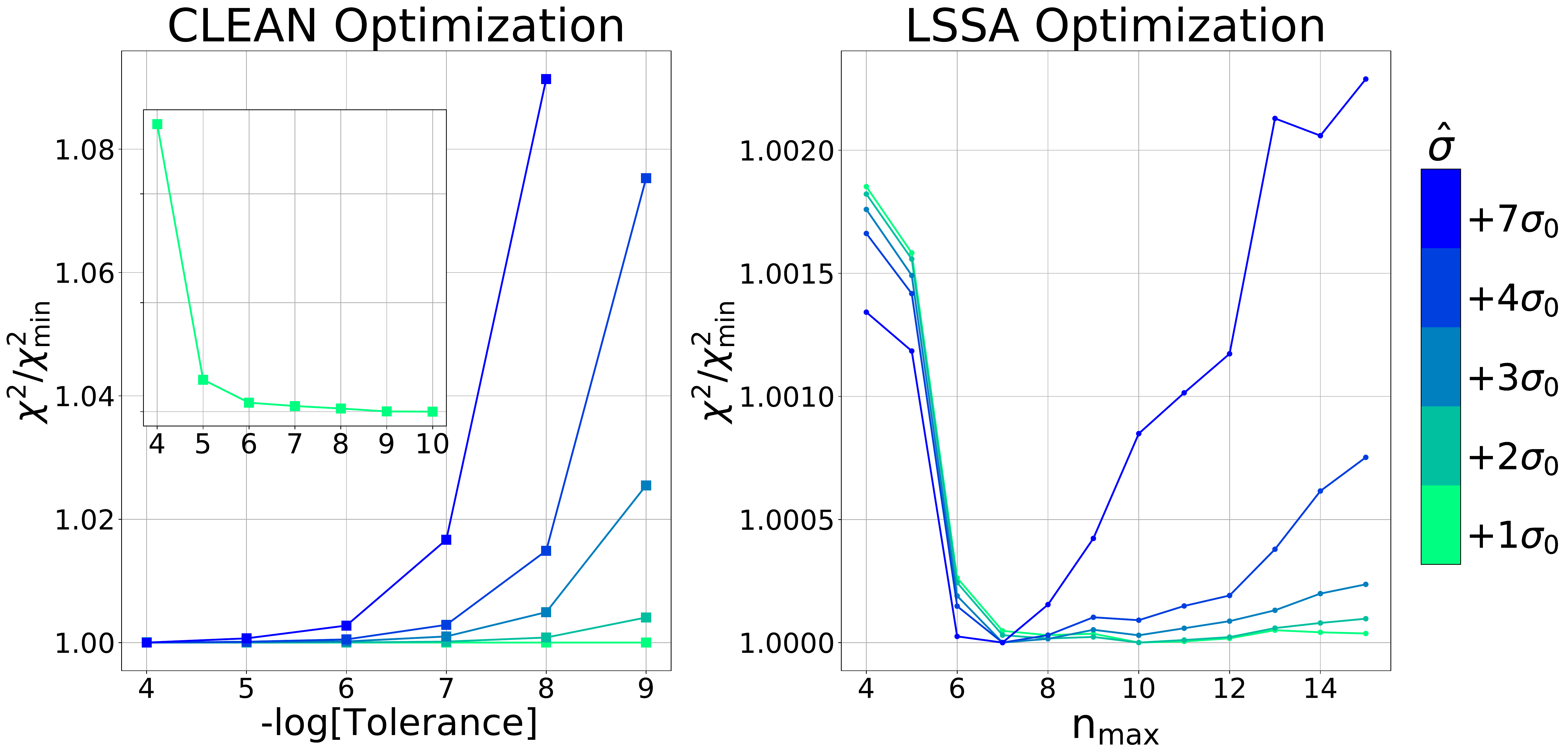}
  \caption{The results of our parameter optimization procedure for CLEAN and LSSA inpainting methods. In the left image fractional increase in $\chi^2$ is plotted as a function of tolerance parameter values (see Section \ref{sssec:CLEAN}). The coloured curves represent different noise levels. As the thermal noise level in the dataset increases the optimal tolerance decreases. The inset provides a closer examination of  of $\chi^2/chi^2_{\rm min}$ for fiducial noise level $\alpha = 1$.  Similarly on the right image, the fractional increase in $\chi^2$ is plotted as a function of $n_{\rm max}$, the number of Fourier components to include in LSSA models. As we increase the thermal noise of the dataset, the optimal number of Fourier components to include in the model decreases.} 
  \label{fig:ParameterOptimization}
\end{figure*}

The implementation of the CLEAN inpainting algorithm in HERA is similar in concept to the algorithm originally introduced in \cite{CLEANHog}. The original algorithm is essentially a deconvolution algorithm for 2D images. The procedure has been slightly modified to fit the needs of inpainting flagged data in the HERA analysis \citep{Parsons2009CLEANvisib, NickKernCalibPartii, HERAH1CData}. For example, the original CLEAN algorithm operates in the image plane whereas the HERA implementation operates in the $\tau$ and $\nu$ domain. More broadly, the original algorithm operates on 2D images whereas the HERA implementation acts independently at each LST taking only the 1D frequency spectrum as input. Since CLEAN operates at each LST independently, LSTs where the entire frequency band are flagged remain flagged. The algorithm works by computing the Fourier transform of the visibilities $\widetilde{V}(\textbf{b}, \tau, t)$ along the frequency axis in accordance with Equation \ref{eq:delay_pspec}. In doing so, the algorithm has an adjustable parameter called the ``zeropad'' parameter, which is the number of bins to zeropad on both sides of the frequency axis. The additional padding around the frequency axis increases the delay space resolution which provides the algorithm with a finer set of discretized $\tau$ modes. The algorithm then iteratively searches and selects the mode $\tau_{i}$ that has the largest amplitude $\widetilde{V}_{\rm max}(\bf{b}, \tau_i, t)$, which is then subtracted from the original quantity, i.e. $\widetilde{V}_1(\bf{b}, \tau_i, t) = \widetilde{V}(\bf{b}, \tau, t) - \widetilde{V}_{\rm max}$. This process is repeated $n$ times until the largest remaining delay modes $\widetilde{V}_n(\bf{b}, \tau_i, t)$ are consistent with the desired tolerance threshold. The tolerance threshold is an adjustable parameter which sets the level at which the algorithm converges. Decreasing this parameter improves performance but is computationally expensive. Another adjustable parameter which determines minimum delay $\tau_{\rm dc}$ is used in estimating the noise, i.e. only delays $\tau > \tau_0$ are used in estimating the noise. This sets a hard cutoff to which modes will be included in the inpainted image. The subtracted delay modes are then used to reconstruct the visibilities in the flagged regions. The CLEAN predictions are referred to the CLEAN model component, whereas the remaining modes are used to construct the CLEAN residual component. 

The accuracy of the CLEAN predictions depend on the input values of the zeropad and tolerance parameters. Thus we need to optimize these parameters. Since the optimal values of the zeropad and tolerance depend on the properties of the dataset, this procedure is repeated for each noise scenario in the simulated data discussed in Section \ref{sssec:P1V_Data}. We find that $\tau_{\rm dc}$ parameter does not dominantly affect the performance and keep the parameter fixed to $\tau_{\rm dc} = 2000$ns unless otherwise noted. To determine the set of optimal parameters of the tolerance and zeropad parameters we compute the sum of the square of the residuals $\epsilon_r$ of Equation \ref{eq:ResidualErrors} between the model visibilities and the true visibilities:
\begin{equation}
    \label{eq:chi2_optimization}
    \chi^2 = \sum_{\rm LST_i, \nu_j }\left[ V_{\rm model}(\rm LST_i, \nu_j ) - V_{\rm true}(\rm LST_i, \nu_j ) \right ]^2
\end{equation}
where we have explicitly made mention to that this sum occurs over all LSTs and frequency channels in the visibilities. Note that it is not necessary to select only the flagged pixels in this sum (i.e. by applying the inverse mask of Equation \ref{eq:loss_function} and \ref{eq:ResidualErrors}), since non-flagged pixels do not contribute to the sum in Equation \ref{eq:chi2_optimization}. The optimal values of these parameters are such that $\chi^2$ in Equation \ref{eq:chi2_optimization} between inpainted predictions relative to the true visibilities are minimized. In Figure \ref{fig:ParameterOptimization} we show the $\chi^2$ for various values of the the tolerance parameter at different thermal noise levels of the dataset. As we increase the noise level, the optimal values the tolerance increase. We find that the behaviour of the zeropad parameter is similar for different thermal noise levels, i.e. increasing the thermal noise of the dataset results necessitates decreasing the value of the zeropad parameter. For the remainder of this paper use CLEAN parameters $\textrm{tol} = 10^{-10}$, $\textrm{zp} = 256$ for the fiducial thermal noise scenario in Section \ref{sssec:P1V_Data} (i.e. $\alpha = 1)$. For $\alpha = 2$ , $3$, $4$ and $7$ we use $\textrm{tol} = 10^{-9}$, $10^{-5}$, $10^{-5}$ , $10^{-4}$. For the zeropad parameter we use $\textrm{zp} = 256$, $256$, $128$, $128$, $64$ respectively.

\subsection{Least Squares Spectral Analysis (LSSA)}
\label{sssec:LSSA}

The HERA implementation of LSSA follows a generalized least squares estimator. It finds a best-fit smooth model derived from the Fourier components of the dataset and uses that model to fill in the flagged regions. This approach is similar in approach to what CLEAN does (see Section \ref{sssec:CLEAN}), except this uses a linear fit rather than the non-linear algorithm of CLEAN. As a result LSSA is computationally less expensive than CLEAN and in principle the error properties are easier to compute. Like the CLEAN algorithm, the code operates at each LST independently, i.e. the best fitting model is derived using the frequency information at each LST. Thus LSSA does not provide a model for LSTs where all frequency channels are flagged.
Consider flagged visibilites at $V(\rm \boldsymbol b, \nu ,t)$ at time $t$, the model for the flagged regions in the visibilities is constructed by expressing $V_{\rm model}(\textbf{b}, \nu, t)$ as a linear combination of the Fourier basis, i.e
\begin{equation}
    \label{eq:Fourier_Basis}
    V_{\rm model}(\textbf{b}, \nu, t) = \sum_{n = -n_{\rm max} }^{n = n_{\rm max} } c_n e^{i\nu n t/ \rm{BW} }
\end{equation}
 where $\rm{BW}$ is the bandwidth of the instrument, $n_{\rm max}$ are the number of user-specified Fourier modes used to model the dataset and $c_n$ are the undetermined coefficients for each Fourier mode. To solve for the coefficients the code uses a linear least squares optimizer, which minimizes the $\chi^2 $residual from Equation \ref{eq:chi2_optimization}.
The solution to Equation \ref{eq:chi2_optimization} is the well known least squares solution.
The best fitting $c_n$ from Equation \ref{eq:chi2_optimization} are then used to construct the model for the visibilities $V_{\rm model}(\textbf{b}, \nu, t)$ in Equation \ref{eq:Fourier_Basis}. The inpainted data are then obtained by replacing $V_{\rm model}(\textbf{b}, \nu, t)$ into the RFI flagged regions of $V_{\rm data}(\textbf{b}, \nu, t)$.

Since the performance of the LSSA algorithm depends on the number of Fourier components $n_{\rm max}$ to include in the model, we need to select $n_{\rm max}$ such that the performance is optimized. We repeat our procedure for each noise scenario in the simulated data discussed in Section \ref{sssec:P1V_Data}. Fewer $n_{\rm max}$ results in a smoother inpainted model while larger values of $n_{\rm max}$ result in producing inpaint models with fine frequency features. For datasets with a greater fraction of flags or larger amplitude of thermal noise, increasing $n_{\rm max}$ too far can hinder the performance due to numerical instabilities. In the case of high percentage of flags, this occurs because there is not enough data to distinguish between the values of the largest Fourier modes. Similarly increasing the thermal noise will expand the error bars of the dataset making it difficult to break the degeneracies between the largest Fourier modes of the LSSA model. In such scenarios performance will be improved with a limited number of modes. We chose $n_{\rm max}$ to strike a balance between goodness of fit and numerical instabilities. To find the optimal value of $n_{\rm max}$, we use the LSSA method to generate models for the RFI flagged regions in the visibilities discussed in Section \ref{sssec:P1V_Data}. We repeat this procedure for multiple values of $n_{\rm max}$ ranging from $n_{\rm max}$ from $2$ to $60$. At each instance we compute the sum of the square of the residuals $\epsilon_r$ of Equation \ref{eq:ResidualErrors} between the model visibilities and the true visibilities, i.e. Equation \ref{eq:chi2_optimization}.
As discussed earlier, note that it is not necessary to select only the flagged pixels in this sum, since non-flagged pixels do not contribute to the sum in Equation \ref{eq:chi2_optimization}. Note that the optimal value of $n_{\rm max}$ depends on which flagged channels we include in our computation of Equation \ref{eq:chi2_optimization}. For example including only the wideband RFI gaps would lead to solutions where fewer modes (smoother functions) are preferred. Conversely applying our optimization to narrowband RFI gaps (for example, the 120MHz - 130MHz in Figure \ref{fig:HERAFlags}) would favour a larger number of Fourier modes. Thus by using all flagged channels in our computation of Equation \ref{eq:chi2_optimization} we strike a balance between models which are best suited for wideband RFI and narrowband RFI.
In Figure \ref{fig:ParameterOptimization} we show the $\chi^2$ as a function of $n_{\rm max}$ for various thermal noise levels. From this we can see that fewer Fourier components lead to better results. We also see that the number of Fourier components to include in the LSSA model decreases with increasing thermal noise. For the remainder of this paper use $n_{\rm max} = 10$ for the fiducial noise scenario, i.e. $\alpha = 1$ in Equation \ref{eq:Radiometer}. For the $\alpha = 2$, $3$, $4$, $7$ thermal noise scenarios we use $n_{\rm max} = 9$, $7$, $7$ ,$6$ respectively.

\begin{figure}
  \includegraphics[width=0.5\textwidth]{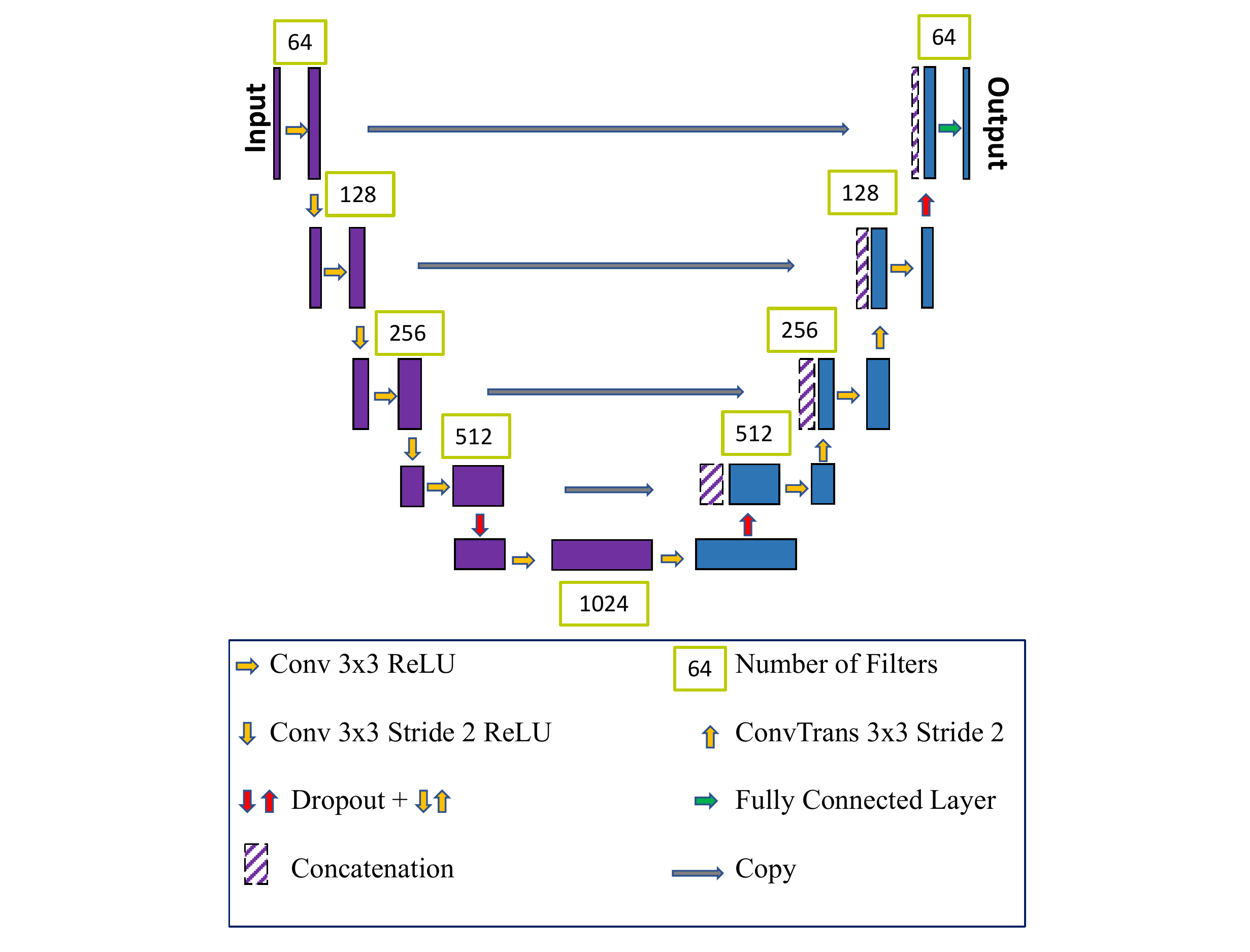}
  \caption{Block diagram showing the U-Paint architecture.  }
  \label{fig:UPaintArch}
\end{figure}

\subsection{Covariance-Based Inpainting (GPR)}
\label{ssec:inv_cov}
A powerful technique for the reconstruction or interpolation of a noisy signal is the Wiener filter \citep{Wiener}, which has a long history in cosmology \citep[e.g.][]{Zaroubi95, Tegmark03}.
A generalization of the Wiener filter is the Gaussian process regression (GPR) formalism \citep{Rybicki92, Rasmussen2006}. Both are, in essence, techniques that down-weight the observed data by its covariance, and then up-weight by the signal covariance.
Recently, GPR has been used in 21\,cm cosmology as a tool for signal separation \citep{Mertens18, Ghosh2020} and for simultaneous filtering and inpainting \citep{Kern2021}.
Following \citet{Kern2021}, the expectation value of the conditioned signal model in a Gaussian process model can be computed as
\begin{align}
\label{eq:gpr}
{\rm E}[s] = C_s(C_s + C_n + C_{\rm other})^{-1}d,
\end{align}
where $d$ is our data vector, ${\rm E}[s]$ is the expectation value of our statistical model for the signal, and $C_s$, $C_n$, and $C_{\rm other}$ are the covariance matrices for the signal, noise and extraneous components of our data model.
This ``best-fit'' also has a covariance given by
\begin{align}
\label{eq:gpr_err}
{\rm Cov}[s] = C_s - C_s(C_s + C_n + C_{\rm other})^{-1}C_s.
\end{align}
Ignoring the $C_{\rm other}$ term in \autoref{eq:gpr}, we see that this indeed simplifies to the standard Wiener filter.
Note that \citet{Kern2021} showed that the GPR foreground subtraction formalism used in 21\,cm cosmology is closely related to the widely studied inverse covariance weighting found in the quadratic estimator literature, in the sense that one first weights the data by its inverse covariance, and the up-weights the residual by a normalization factor. More generally, typical applications of GPR involve fitting for the hyperparameters of analytic covariance functions, but at the end of the day, GPR is simply an inverse covariance weighting, as shown above. Further note that any covariance function can be implemented within the GPR framework discussed above \citep[e.g.][]{Ghosh2020}.

In this work, we adopt a simple squared-exponential covariance function for modeling the 21\,cm foregrounds, and a diagonal matrix for modeling the (uncorrelated) thermal noise. The hyperparameters of these covariances (e.g. the squared-exponential length scale and the noise variance) were set manually via inspection of the data: although one could choose to regress for these automatically on the data, given our understanding of the datasets at-hand we found that manual selection yielded similar results.

Another recent example of covariance-based modeling for 21\,cm is the DAYENU formalism of \citet{EW21}.
Fundamentally, DAYENU is an inverse-covariance technique that explicitly assumes a Sinc model for the frequency-frequency covariance of the visibilities. Note that DAYENU was designed as a filter to remove foregrounds, however, the construction of the filter to remove this signal is similar to that of \autoref{eq:gpr}.
In fact, although not explicitly shown in \citet{EW21}, one can see that DAYENU is exactly the same as \autoref{eq:gpr} in the case of a signal covariance that is the identity matrix, and a noise covariance that is a sinc function.
The set of vectors that diagonalize this sinc covariance are the discrete prolate spheroidal sequences (DPSS), which have a long history in signal processing as the solution to the spectral concentration problem \citep{DPSSintro}.

\subsection{DPSS Least Squares (DPSS-LS)}
The LSSA technique discussed in the previous section can be generalized to model functions (instead of just fourier components). In general, we can model the visibility data at a single time as 
\begin{equation}\label{eq:SMOOTH_FIT}
V_\text{model}(\text{LST}_i, \nu_j) = \sum_\alpha A_\alpha(\text{LST}_i) u_\alpha(\text{LST}_i, \nu_j)
\end{equation}
where $u_\alpha$ are a set of vectors that ideally span all possible foreground shapes while having minimal overlap with modes outside the wedge. Since foregrounds within the wedge are heavily ``band-limited'' -- are ideally only contained within a compact range of delays, sets of functions whose Fourier transforms maximize power within a band-limited region are are ideal for describing these foregrounds. The Discrete Prolate Spheroidal Sequences (DPSS) \citep{DPSSintro} maximize the ratio of power within some bandlimited region $B_\tau$ to the total power of the sequence and are thus an ideal basis for per-baseline modeling of the wedge. \citet{EW21} applied these sequences to modeling and filtering foregrounds with the DAYENU technique in which the covariance matrix of foregrounds is approximated as a Sinc matrix which is diagonalized by DPSS modes or DAYENUREST which performs linear least-squares inpainting.

Although the DAYENU (i.e. DPSS) formalism presented in \citet{EW21} and discussed above is presented as a covariance-based technique similar to the Wiener filter and GPR, there are other ways to use the DPSS vectors for data modeling and inpainting. The DAYENUREST variant presented in \citet{EW21} does just this, and instead of inpainting via \autoref{eq:gpr}, it uses the DPSS vectors as a basis-set for performing least-squares fitting in the visibility. In this sense, the DAYENUREST (or DPSS least squares) is more akin to the LSSA formalism discussed above, except with a DPSS basis set instead of discrete Fourier modes. Hereafter, when we refer to ``DPSS'' in the paper we refer specifically to the DPSS least squares technique, which is distinctly separate from the pure covariance-based inpainting techniques like GPR.
Similar to LSSA we must specify how many modes to include in our DPSS basis-set. To do this, one specifies the parameter $\tau_{\rm dc}$ which determines the the finest spectral scale that DPSS inpaints over, i.e. $1/\tau_{\rm dc}$. Increasing $\tau_{\rm dc}$ results in capturing finer frequency structures while decreasing $\tau_{\rm dc}$ results in modeling only the smoothest frequency structures. Thus the maximum RFI gap that is inpainted is proportional to $1/\tau_{\rm dc}$. Similar to selecting $n_{\rm max}$ in Section \ref{sssec:LSSA}, our selection of $\tau_{\rm dc}$ has consequences for the performance of the model in narrowband relative to wideband RFI. For example, increasing $\tau_{\rm dc}$ results in inpaint models which can account for fine frequency structure, which optimizes the performance for narrowband RFI. Conversely, this means that there is a maximum RFI gap size $1/\tau_{\rm dc}$ for which we can inpaint over which reduces performance in wideband RFI gaps. In this paper we use $\tau_{\rm dc} = 1000$ns. This makes our DPSS technique optimized at inpainting intermittent (i.e. narrowband) RFI and introducing a maximum gap size of 1/$\tau_{dc} = 0.5$MHz. Since this technique is similar to that of LSSA, and because our parameter choices for DPSS and LSSA optimize performance for different RFI properties, our analysis essentially brackets the range of performance for DPSS and LSSA techniques. 

 \begin{figure*}
  \includegraphics[width=0.99\textwidth]{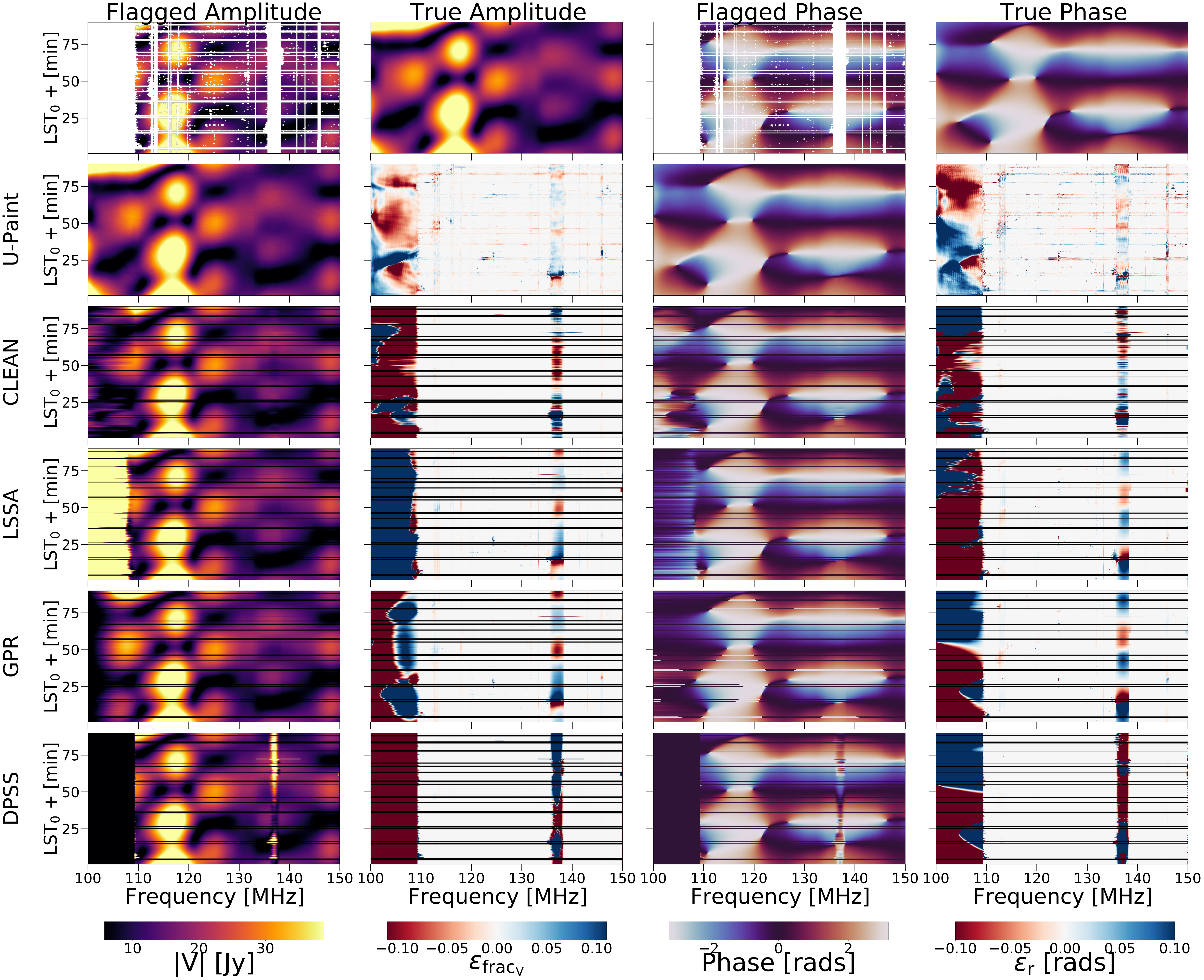}
  \caption{First row: The amplitude and phase components of the RFI flagged visibilities are shown in the first and third column. In the second and fourth column are the amplitude and phase component of the true visibilities. The visibilities are simulated (see Section \ref{sssec:P1V_Data}). Second through fifth rows: in each row we show the amplitude and phase components of the RFI flagged visibilities but with the inpaint models filled into the RFI gaps. Each subsequent row correspond to U-Paint, CLEAN, LSSA, GPR, and DPSS inpainting methods.  In the second and fourth column of each row we show the fractional error of the model amplitude and the residuals of the model phase (see Section \ref{sec:VisibilityErrorQuantification}). }
  \label{fig:5pt2AmplitudePredictionsAll}
\end{figure*}

\subsection{U-Paint Architecture}
\label{sssec:Unets}
Our desired network configuration is one which is capable at making precise predictions of the data in flagged regions using the unflagged features in the visibilities. To do this we use a U-net architecture, introduced by \cite{UNETRonennberger} which have been shown to be robust for these type problems \citep{IsenseeUNet2018}. Our U-Net construction closely follows the architecture of \cite{UNETRonennberger} and \cite{SamWedge}. We show the schematic of our network in Figure \ref{fig:UPaintArch}. Starting from the input of Figure \ref{fig:UPaintArch}, we input images of size 512$\times$512. As discussed in Section \ref{sssec:P1V_Data}, we use data from antennas $(84,85)$ and $(0,1)$ to perform our analysis. Thus all data from these antennas are removed before training. As discussed in Section \ref{sssec:P1V_Data} the HERA visibilities are measures of 1024 frequency channels over $4000$ time integrations (i.e $\times N_{\rm LSTs}$). Thus we divide the total HERA visibilities into input visibilities of size $512\times512$ corresponding to 90min of data and a band width of $50$MHz. Thus the frequency band is split into two sections $100$MHz-$150$MHz and $150$MHz-$200$MHz at 90min observation intervals. Our motivation for selecting visibility sizes of 512$\times$512 is to establish a balance between two considerations: we need to divide the visibilites enough times to generate a large enough dataset for training and while simultaneously allowing a large enough image to allow the network to recognize typical features in HERA visibilities. Segmenting the data into too small a size will obscure the larger features in the visibilities. Conversely, making the image size too large will reduce the amount of images in our training set. Note that we find that the performance of the network is similar when using image sizes of 256x256; however, we find that the performance of the network is decreased below this threshold. Each visibility image is then split into 5 input channels\footnote{In this subsection, ``channels'' refers to the inputs to convolutional layers and not frequency channels. Outside of this subsection, channels refers to frequency channels.} for the initial convolutional layer. Thus the input has shape $512 \times 512  \times 5$. Our input channels are as follows: in channels 1 \& 2 we input the real and imaginary component of the visibilities, respectively, defined in Equation \ref{eq:visibilityDefinition} where the flagged regions of the real and imaginary component of the visibilities have been set to 0. In channel 3 we input the flags, which are a binarized 512 $\times$ 512 map where a 0 pixel represents an unflagged region in the visibilities and 1 represents a flagged region in the visibilities. In order to ensure continuity at the boundary between flagged regions and the unflagged regions, i.e. between our inpainted predictions and the existing visibilities, we extend the flagged regions by two adjacent pixels along both axes (i.e. in LST and $\nu$). This encourages the network's model of the visibilities to be consistent with the existing information in the unflagged regions. In channels 4 \& 5 we input the real and imaginary component of $\widetilde{V}(\textbf{b}, \tau, t)$, i.e. Equation \ref{eq:delay_pspec} is applied to the visibilities $V(\textbf{b}, \tau, t)$ within channels 1 \& 2 respectively. This is done to encourage the network to take advantage of the delay information. The reason this is effective is because our data is structured in the delay domain: high power at low delays due to the foregrounds and then lower power at high delays due to noise. 

Referring again to architecture of the network in Figure \ref{fig:UPaintArch}, the objective of left branch of the U-net is to capture context of the images and propagate them downward through each level. We choose convolutional kernels of size $(2 \times 2 )$ which gives us a reasonable balance between the spatial resolutions and context for the features comprising the image. At each level we use a ``ReLU'' activation function. As the input data is propagated through each level, the network increasingly forms an abstraction of the elements in the image. The bottom of the U-net can be interpreted as a classification type step, i.e., at this stage the network has understood the various elements in the image and has formed an abstract classification of these items. The objective of the right side of the U-net is to use the abstract classification of the items in the image to make predictions of the data in flagged regions of the input dataset. To do this the network uses a convolutional layer which upscales the size of each image. Throughout this process the network has lost all context about the superficial placement of these features. To re-introduce the necessary superficial context to each level on the right side of the U-net, skip connections between the levels on the left branch of the U-net and right branch of the U-net are formed. The image on the left hand side of the U-net is combined with the corresponding level on the right hand side through concatenation. The output at the right of Figure \ref{fig:UPaintArch} has shape $512\times512$ and contains the network's model for the flagged regions. We extract the network predictions for the flagged regions of the visibilities and insert them into the corresponding flagged regions of the original flagged data set. In other words, we discard the network's predictions for the data in unflagged regions.

To compare the training set to the labels, we define difference between the model visibilities $V_{\rm model}(\nu, t)$ and labels $V_{\rm true}(\nu, t)$ as $\Delta = V_{\rm model}(\nu, t) - V_{\rm true}(\nu, t)$. We use a loss function
\begin{eqnarray}
    \label{eq:loss_function}
    \chi^2 = \sum_{n} \left [ \left(\mathds{1} - M(\nu,t) \right )  \Delta \right]^\dagger   \cdot \left [ \left(\mathds{1} - M(\nu,t) \right )  \Delta \right]
\end{eqnarray}
where the sum is over $n$, the number of images in the batch. The $\dagger$ refers to complex conjugation and a transpose. The term $\mathds{1} - M(\nu,t)$ essentially inverts the flags, i.e. the unflagged regions are 0 and the flagged regions are 1. The inverse flags prevent non-flagged regions from contributing to the loss. This is done to encourage the network to focus on learning the features of the flagged regions, which speeds up our training process. 

We use $\sim350$ images from the the simulated visibilities discussed in Section \ref{sssec:P1V_Data} as part of our training set, and a test set of $35$, with a batch size of $12$. The network is trained for $80$ epochs and a learning rate of $lr = 10^{-4}$ using an \texttt{Adam} optimizer. 

\section{Inpaint Models}
\label{sec:InpaintModelsIntro}

 In this section we use the inpainting methods to make predictions for the RFI corrupted simulated visibilities from Section \ref{sssec:P1V_Data}. We also provide a high level qualitative overview of the inpainted models in their amplitude and phase components. In Figure \ref{fig:5pt2AmplitudePredictionsAll} we show sample inpaint predictions for the amplitude and phase of the RFI corrupted visibilities. The upper left panel of Figure \ref{fig:5pt2AmplitudePredictionsAll} corresponds to the flagged visibilities while the top of the second column corresponds to the true visibilities. The first column in each subsequent row corresponds to visibilities where the inpaint models have been replaced in the RFI flagged regions. The first row corresponds to U-Paint models, the second row corresponds to CLEAN models, the third row corresponds to LSSA models and the final two rows correspond to GPR and DPSS models respectively. The attributes of the  predictions shown in this image are characteristic of the models for each inpainting method. By visual inspection we can see that the U-Paint network has learned to assimilate the features in the amplitude and phase into the RFI corrupted regions, and thus it is apparent that the network is capable of reproducing the features of the true visibilities in the RFI corrupted regions. Another distinguishing feature of the network predictions are that the network organically inpaints over LSTs that do not contain any frequency information. In contrast to the other inpaint algorithms which do not naturally provide predictions for these LSTs, U-Paint can take advantage of all the information of the visibilities. This highlights U-Paint's ability to extrapolate data to LSTs in which there are none. Currently LSTs without any frequency information are not used as part of HERA's data analysis pipeline; however, in the future one may be able to take advantage of these LSTs either from the analysis perspective or simply to avoid discontinuities in the data. We can also see that all inpainting methods do a reasonable job at filling in the narrowband RFI portion of the visibilities making it difficult to discern between the true visibilities and the inpaint models. In contrast, regions where wideband RFI has been replaced with inpaint models are still obvious. Referring to the 2MHz RFI gap at 136MHz, we can see that wideband RFI is still easily identifiable in the model visibilities of each inpainting technique. There appear to be remaining artifacts in the wideband RFI regions which make the characteristics of the inpaint models are apparent. Referring to the top row, can see that U-Paint produces models with a speckled structure in frequency while CLEAN, LSSA and GPR models tend to be smoother in the frequency domain. DPSS models don't entirely fill in the wideband RFI gap at 136MHz. As discussed in Section \ref{sec:InpaintingTechniques}, this is due to our choice of delay cut parameter $\tau_{\rm dc}$. The maximum RFI gap that is inpainted is proportional to $1/\tau_{\rm dc}$. Since we are using $\tau_{\rm dc} = 1000$ns, then we are limited to RFI gaps larger than 1/$\tau_{dc} = 0.5$MHz. Unless otherwise stated we do not include DPSS in our error characterisation for wideband RFI. 
 In the third column of Figure \ref{fig:5pt2AmplitudePredictionsAll} we show the phase component of the inpaint predictions. The second through fifth rows again correspond to U-Paint, CLEAN, LSSA, GPR, and DPSS models respectively. We can see that the inpaint models capture the structure of the phase component. As was the case with the amplitude component, regions of inpainted narrowband RFI appear to be seamlessly integrated with the rest of the visibilities while inpainted wideband regions appear to have artifacts. 
 
 In the following sections we build a quantitative perspective on the performance of each inpainting technique. In the next section we discuss our methodology in quantifying the error characteristics of the inpaint models.

\section{Statistical Analysis Methodology}
\label{sec:VisibilityErrorQuantification}



We quantify the errors in inpainted predictions relative to the true visibilities by computing the residuals, fractional errors and a modified version of the fractional errors. We use the same metrics to quantify the errors in the model power spectra relative to the true power spectra. The residuals between the inpainted visibilites and the true visibilities are computed as 
\begin{equation}
    \label{eq:ResidualErrors}
    \epsilon^{ \rm V }_{\rm r}=  \left [ \mathds{1} - M(\nu, t)\right ] \cdot \left (V_{\rm model} - V_{\rm true} \right) ,
\end{equation}
  where $M(\nu, t)$ are the flags, $V_{\rm model}$ are the flagged visibilities where the inpainted models have been placed into the flagged regions and $V_{\rm true}$ are the true visibilities (i.e. without any flags). The term $1 - M(\nu, t)$ essentially inverts the flags i.e. 1 is a flagged region and 0 signifies unflagged. This is done so that only flagged regions enter the analysis. As discussed in Sections \ref{sssec:CLEAN}, \ref{sssec:LSSA}, and \ref{ssec:inv_cov}, CLEAN, LSSA, GPR, and DPSS operate at each LST independently and thus do not inpaint on LSTs where the entire frequency band are flagged. These LSTs are not used in our error characterization analysis even for inpainting methods which do inpaint on these LSTs, i.e. U-Paint. 
Note that the residuals defined by Equations \ref{eq:ResidualErrors} constitute individual error realisations. In Section \ref{sec:VisibilityErrorQuantification} we model the distribution of error realisations to compute the actual error. Using $\epsilon^{\rm V}_{\rm r}$ we can define the fractional error $\epsilon_{\rm frac}$:
\begin{equation}
    \label{eq:frac_errors_def}
    \epsilon^{ \rm V }_{\rm frac} =  \frac{ \epsilon^{\rm V}_{\rm r} }{V_{\rm true}}.
\end{equation}

Since the visibilities are complex, they can be split into real and imaginary components, or amplitude and phase. Within the context of error quantification, Equations \ref{eq:ResidualErrors} and \ref{eq:frac_errors_def} can be applied to the real, imaginary, and amplitude components of the visibilities. However since the phase of the visibilities are periodic, quantifying the errors using the fractional errors defined in Equation \ref{eq:frac_errors_def} becomes meaningless. To quantify the errors for the phase component of the visibilities we use a modified version of the residuals of Equation \ref{eq:ResidualErrors}. The phase values of the inpainted models $\phi_{\rm model}$ and ground truth $\phi_{\rm true}$ are mapped from their native range $[-\pi,\pi]$ to $[0,2\pi]$. The residuals $\Delta \phi = \phi_{\rm model} - \phi_{\rm true}$ are then computed. Since the sign of the phase error does not directly indicate the severity of the error, i.e, a phase error of $+\Delta \phi$ is the same ``angular distance'' from the true value as phase error $-\Delta \phi$, we define the absolute residual phase error $\epsilon_\phi$ as  
\begin{equation}
\label{eq:CircError}
    \epsilon_\phi = \textrm{min} \left ( \left | 2\pi - (\phi_{\rm model} - \phi_{\rm true} )   \right | , \left | \phi_{\rm model} - \phi_{\rm true}    \right | \right ) .
\end{equation}
Therefore we can interpret $\epsilon_\phi$ to be the smallest angle from $\phi_{\rm true}$. In Sections \ref{sec:BigSectionErrorQuantification} and \ref{sec:PowerSpectrumErrorCharacterization} we use these metrics as tools to describe the errors in the model visibilities and power spectra. 

To perform our analysis we construct a sample set of RFI flagged channels using all flagged channels  between $\nu = 110$MHz and $\nu = 174$MHz (see Section \ref{sssec:RFI_introduction} for details). We exclude LSTs in which all frequency channels are flagged from our analysis. As discussed in Section \ref{sssec:P1V_Data} we consider only the shortest baselines (i.e. antennas separated by 14.7m) in this work. We find that our results do not depend on the specific antennas used to form the 14.7m baseline. Thus without loss of generality we perform our analysis using the antennas $(0,1)$ and $(84,85)$ for strictly east-west baselines, including multiple linear polarisations ( EE and NN ). We have repeated our subsequent analyses for redundant baselines using other antenna pairs and have found no significant differences in our qualitative or quantitative results. With the restrictions above, this leads to a sample set of $10^4$ flagged channels. Using this sample set, we construct the empirical error distribution. We model the empirical error distribution with seven main classes of model probability density functions, which along with their sub-classes, encompass a flexible range of probability profiles. They include the gamma, log normal, skew Cauchy (see \cite{SkewCauchyDistributionGupta}), t, skew normal, generalized normal, skew Laplace distributions. These distribution functions comprise a family of distributions in which we find more familiar probability profiles as special cases. 
We then compare the empirical distribution to $p_{\rm best}$ using the Kolmogorov-Smirnov (KS) test introduced in \cite{KSTestOG}. 
In the following Sections we apply these metrics to the inpainted predictions of U-Paint, CLEAN, LSSA, GPR, and DPSS.

\begin{table}
\caption{Summary of key error metrics for the amplitude component of simulated visibilities.  } 
\label{tab:SimDataVis}
\begin{center}
\begin{tabular}{|c|c|c|c|c|} 
\hline

Error & $\overline{\sigma}_{{\epsilon}_{\rm frac}}$ & $\overline{\sigma}_{{\epsilon}_{\rm frac}}$  & $\overline{\mu}_{{\epsilon}_{\rm frac}}$ & $\overline{\mu}_{{\epsilon}_{\rm frac}}$\\ 
  RFI       &   Narrowband   &   All   & Narrowband & All          \\
\hline\hline
U-Paint & 5.5\% & 5.97\% &  -0.025\% &  0.05\%\\
\hline
CLEAN & 3\% & 9.7\% &  0.07\% &  0.265\% \\ 
\hline
LSSA & 1.69\% & 2.82\% &  0.05\% &  -0.16\%\\ 
\hline
GPR & 3.09\% & 3.5\% &  -0.08\% &  -0.044\%\\ 
\hline
DPSS & 1.52\% & - &  0.013\% &  \\ 
\end{tabular}
\end{center}
\end{table}

\section{Inpaint Error Quantification in the Visibilities of Simulated Data}
\label{sec:BigSectionErrorQuantification}
In Section \ref{sec:InpaintModelsIntro} we discussed the qualitative features of the inpaint models. We now examine the quantitative aspects of their errors using the metrics from Section \ref{sec:VisibilityErrorQuantification}. Since the visibilities are complex valued, they can be expressed in terms of amplitude and phase components. In Section \ref{ssec:VisibilityAnalysis} we apply our analysis to both components of the visibilities. In Section \ref{sssec:ThermNoiseChap5} we discuss the impact that increased thermal noise have on the inpaint models. 


 \begin{figure}
  \includegraphics[width=0.45\textwidth]{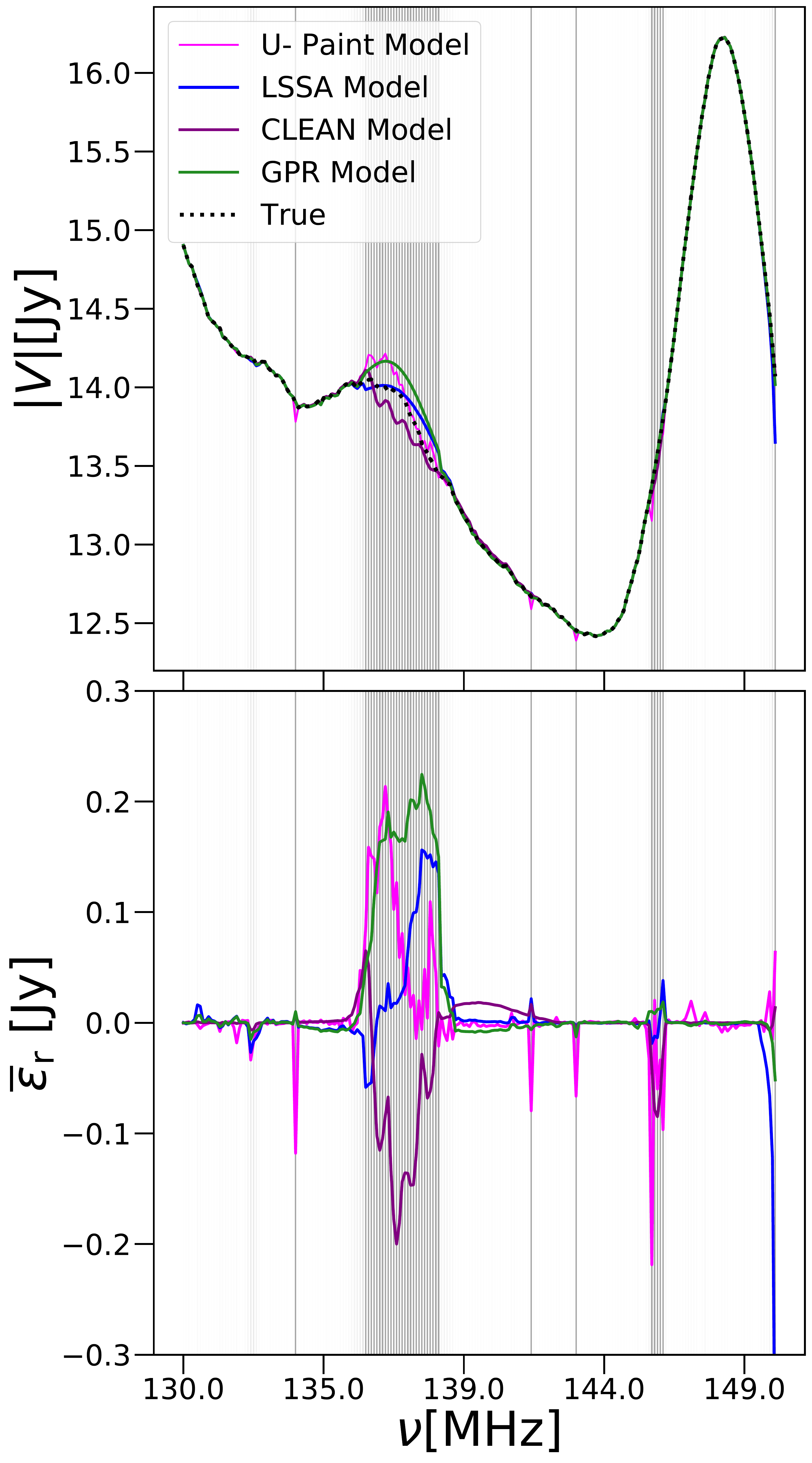}
  \caption{Top: LST averaged inpaint model visibilities. The true visibilities are shown with the dotted black curve. The vertical shaded regions correspond  to the RFI flagged channels. The amount of shade is proportional to the frequency in which those channels are flagged. Thus the Wideband ORBCOMM feature is darkest since it is always flagged. Note that the inpaint models are only filled into RFI gaps, and so the inpaint models only deviate from the true visibilities in shaded regions. The orange curve corresponds to U-Paint, the yellow curve to LSSA, purple curve to CLEAN and blue curve to GPR. DPSS models are not shown since we feature the wideband feature in this image (see Section \ref{sec:InpaintingTechniques}). Bottom: The residuals between inpaint models and the true visibilities.  }
  \label{fig:5pt2Vis_AvgCrossSection_pred0}
\end{figure}

\subsection{Error Characterisation}
\label{sssec:CLEANandLSSA_VisibilityAnalysis}
\label{ssec:VisibilityAnalysis}
 \begin{figure*}
  \includegraphics[width=0.9\textwidth]{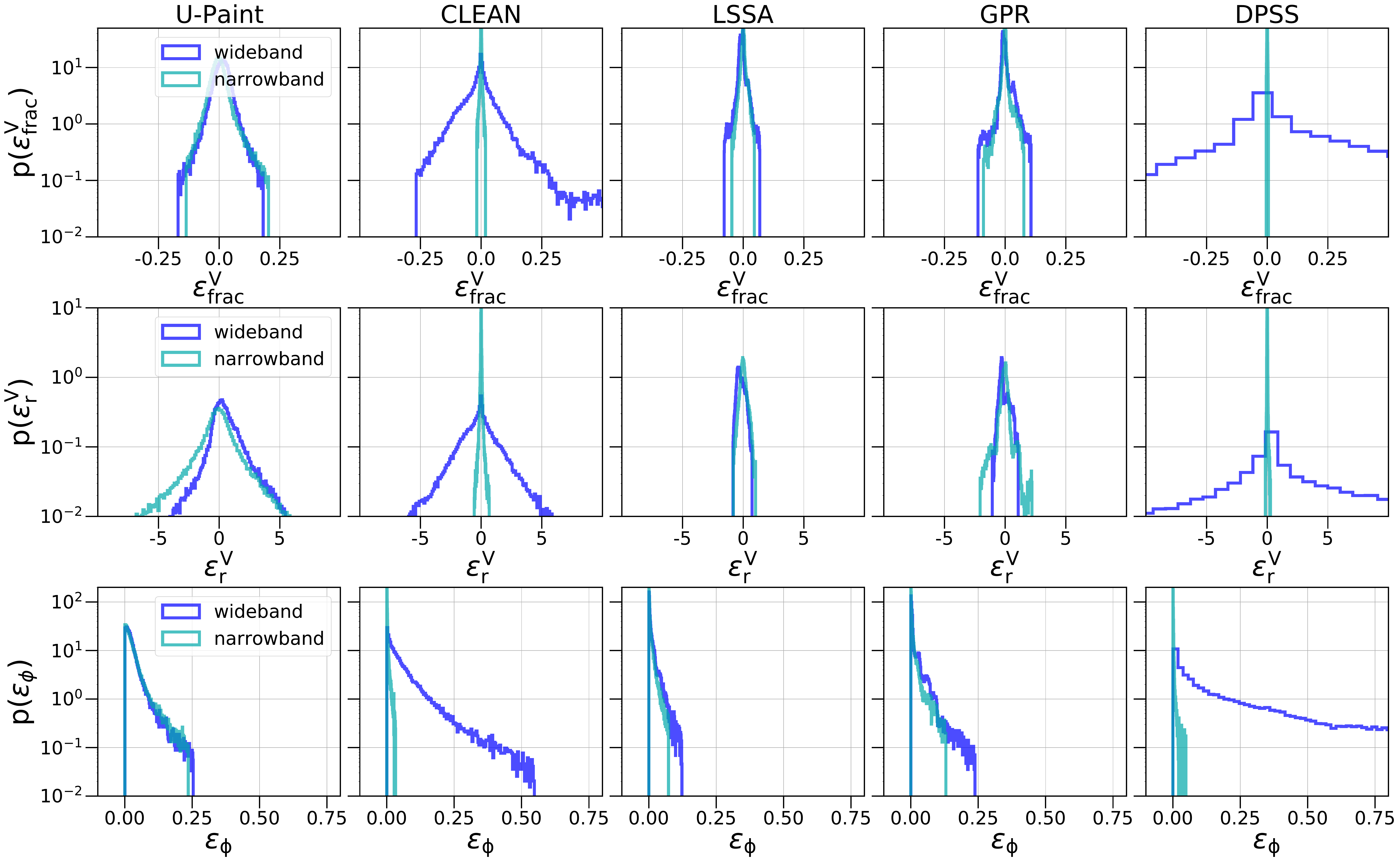}
  \caption{Top row: probability distribution of the fractional errors $p(\epsilon^{\rm V}_{\rm frac})$ in the amplitude of the inpainted model visibilities. Second row: residuals in the inpainted model amplitudes $p(\epsilon^{\rm V}_{\rm r})$. Third row: residuals of the phase component of the inpaint models $p(\epsilon_{\rm phi})$. The blue curves corresponds to when only wideband RFI is used to construct the samples while the teal curve corresponds to samples constructed using only narrowband RFI. All inpaint methods are applied to the simulated visabilities discussed in Section  \ref{sssec:P1V_Data}.}.
  \label{fig:5pt3HistogramMegaPlot}
\end{figure*}
In the second column of Figure \ref{fig:5pt2AmplitudePredictionsAll} we show example fractional errors of the amplitude of the inpaint models. With this metric the errors are normalized by the amplitude of the true visibilities allowing us to ascertain the performance independent of the brightness of the visibilities. Referring to the second row of the second column in Figure \ref{fig:5pt2AmplitudePredictionsAll}, we find that the mean fractional error in the amplitude of U-Paint models is $\mu_{\epsilon_{\rm frac}}  = 0.058\%$ and standard deviation $\sigma_{\epsilon_{\rm frac}} = 5.5\%$ \footnote{Note that in this assessment we are not including the model predictions at $\nu < 110$MHz.}. Thus the fluctuation in performance is $5.5\%$. We also find that $\sigma_{\epsilon_{\rm frac}}$ is consistent throughout the various types of RFI, i.e in wideband and narrowband RFI. We also find that U-Paint has similar performance in LSTs which are entirely flagged. In rows three through six of Figure \ref{fig:5pt2AmplitudePredictionsAll} we show the fractional error in the amplitude of the inpainted models for  CLEAN, LSSA, GPR, and DPSS algorithms. Immediately clear from the fractional errors of the visibility amplitudes are that CLEAN, LSSA, GPR, and DPSS models are more accurate in the narrowband RFI regions as compared to the wideband RFI. The standard deviation of the fractional errors of the inpainted models in narrowband RFI are $\sigma_{\epsilon_{\rm frac}}$ are smallest for DPSS at $1.52\%$ and LSSA at $1.69\%$ followed by CLEAN and GPR at $3.0\%$ and $3.09\%$ respectively. When we include flagged channels above 110MHz, the error fluctuations $\sigma_{\epsilon_{\rm frac}}$ increase. This is due to the inclusion of wideband RFI gaps where the fractional errors are larger. When including all flagged channels above $110$MHz we find that LSSA produces the smallest fluctuations at $2.8\%$ followed by GPR at $3.5\%$, U-Paint at $5.95\%$, CLEAN at $9.7\%$ and DPSS at $10.3\%$ Recall that for DPSS our choice of parameters leads to model limitations on large RFI gaps and thus we do not include DPSS in our error characterisation for wideband RFI. In Table \ref{tab:SimDataVis} we provide a summary of these quantitative results. 

Another distinctive characteristic of the amplitude in U-Paint models are that they contain fine frequency structure. In the top panel of Figure \ref{fig:5pt2Vis_AvgCrossSection_pred0} we show the amplitude of the visibilities as a function of $\nu$ averaged over $512$ time integrations. The dotted black line corresponds to the true visibilities, while the solid colored curves corresponds to the inpaint models. The amount of grey shading represents the average flag occupancy of each frequency bin. In the wideband RFI gap we can closely examine the features of each inpaint model. In the lower panel of Figure \ref{fig:5pt2Vis_AvgCrossSection_pred0} we can see the spectral structure in the residuals between the inpaint model and true visibilities. Note the rapid fluctuating components in the U-Paint predictions as compared to the smoother true visibilities.

In Figure \ref{fig:5pt3HistogramMegaPlot} we show the probability distributions of the fractional errors $p(\epsilon_{\rm frac})$ in the inpainted models. Since the performance and errors depend on the nature of the RFI, we separate our analysis into frequency channels which are dominated by narrowband RFI and frequency channels which are dominated by wideband RFI. For the narrowband RFI we construct a sample set using all flagged pixels from frequency channels $110$MHz to $136$MHz, where these bounds exclude the wideband features found below $110$MHz and above $136$MHz. This leads to a sample size of $\sim 52000$ pixels. For the wideband regions, we isolate the 20 frequency channels corresponding to the the ORBCOMM RFI feature at $136$MHz. This leads to a similar sample size of $54000$ pixels. In the top row Figure \ref{fig:5pt3HistogramMegaPlot} we show the probability density functions of the fractional error $p(\epsilon^{\rm V}_{\rm frac})$ (Equation \ref{eq:frac_errors_def}) for the amplitude of the inpainted models in narrowband and wideband RFI regions. The blue curves correspond to the probability distribution constructed using only the wideband RFI samples, while the teal curve corresponds to the probability distribution constructed using only the narrowband RFI samples. For the sake of visualization, we display up to the $99.9$ percentile of errors along the horizontal axis.  
By qualitatively comparing the maximum range of the teal curve to the blue curve in all five panels of the first row in Figure \ref{fig:5pt3HistogramMegaPlot} we can see that the U-Paint, LSSA, and GPR performances are more consistent across wideband and narrowband RFI regions as compared to CLEAN and DPSS which perform significantly better with narrowband RFI. Note that DPSS does not inpaint over a 2MHz gap given our parameter choices in Section \ref{sec:InpaintingTechniques}. We can also see that the maximum range of fractional errors for narrowband RFI is smallest for DPSS inpainting methods and largest for U-Paint. Conversely, for wideband RFI, LSSA and GPR produce the smallest range of fractional errors. Another feature of the distribution of fractional errors $\epsilon^{\rm V}_{\rm frac}$ for wideband RFI using CLEAN is the positive skew, i.e. a disproportionate amount of probability mass is contained in $p(\epsilon^{\rm V}_{\rm frac}) > 0$. With this exception of this distribution, we find that generalized normal distributions is an optimal probability distribution profile to model the empirical distributions $p(\epsilon^{\rm V}_{\rm frac})$ for each RFI scenario in Figure \ref{fig:5pt3HistogramMegaPlot}.

To establish the range of absolute temperature errors introduced into the analysis, we now examine the distribution of residuals $p(\epsilon^{\rm V}_{\rm r})$ in $\left |V\right|$. The distribution of residuals are shown in the second row of Figure \ref{fig:5pt3HistogramMegaPlot}. Many of the qualitative features in $p(\epsilon^{\rm V}_{\rm r})$ are similar to the distributions of fractional errors from above. For example, the distribution of residuals in U-Paint, GPR, and LSSA inpainting methods are less sensitive to the type of RFI, i.e. narrowband and wideband. By comparing the maximum range of residuals for narrowband RFI for each inpainting technique we again come to the same conclusion as above: DPSS produces the smallest residuals, followed by CLEAN. Similarly, when for wideband RFI, LSSA and GPR produce the smallest residuals. 

We now discuss the distribution of errors in the phase components of the visibilities. Referring to the fourth column of Figure \ref{fig:5pt2AmplitudePredictionsAll} we show the residuals between the model phase and true phase. We see that with the exception of the wideband models for DPSS inpaint methods, all of the residuals fall between $|\epsilon_{\rm r}^{\phi}| < 0.1$rads. The largest residuals are sourced from wideband RFI regions. In the bottom row of Figure \ref{fig:5pt3HistogramMegaPlot} we show the corresponding distributions of the residual phase errors $\epsilon_\phi$ as defined in Equation \ref{eq:CircError}. Recall that the errors $\epsilon_\phi$ are bounded between $\epsilon_\pi = 0$ and $\epsilon_\phi = \pi$. We find that the errors in the phase component $\epsilon_\phi$ of the inpainted models are all characterized by the same type of probability distribution profile, the lognormal probability function. Similar to our descriptions of $p(\epsilon^{\rm V}_{\rm r})$ and $p(\epsilon^{\rm V}_{\rm frac})$, we find that CLEAN and DPSS models provide the most accurate description of the phase in narrowband RFI regions and LSSA providing the best description of the phase in wideband RFI. Relative to DPSS and CLEAN inpainting methods, we again find U-Paint, GPR and LSSA have consistent performance in the phase component for the narrowband and wideband RFI.

\subsection{Thermal Noise}
\label{sssec:ThermNoiseChap5}
  

 \begin{figure}
  \includegraphics[width=0.51\textwidth]{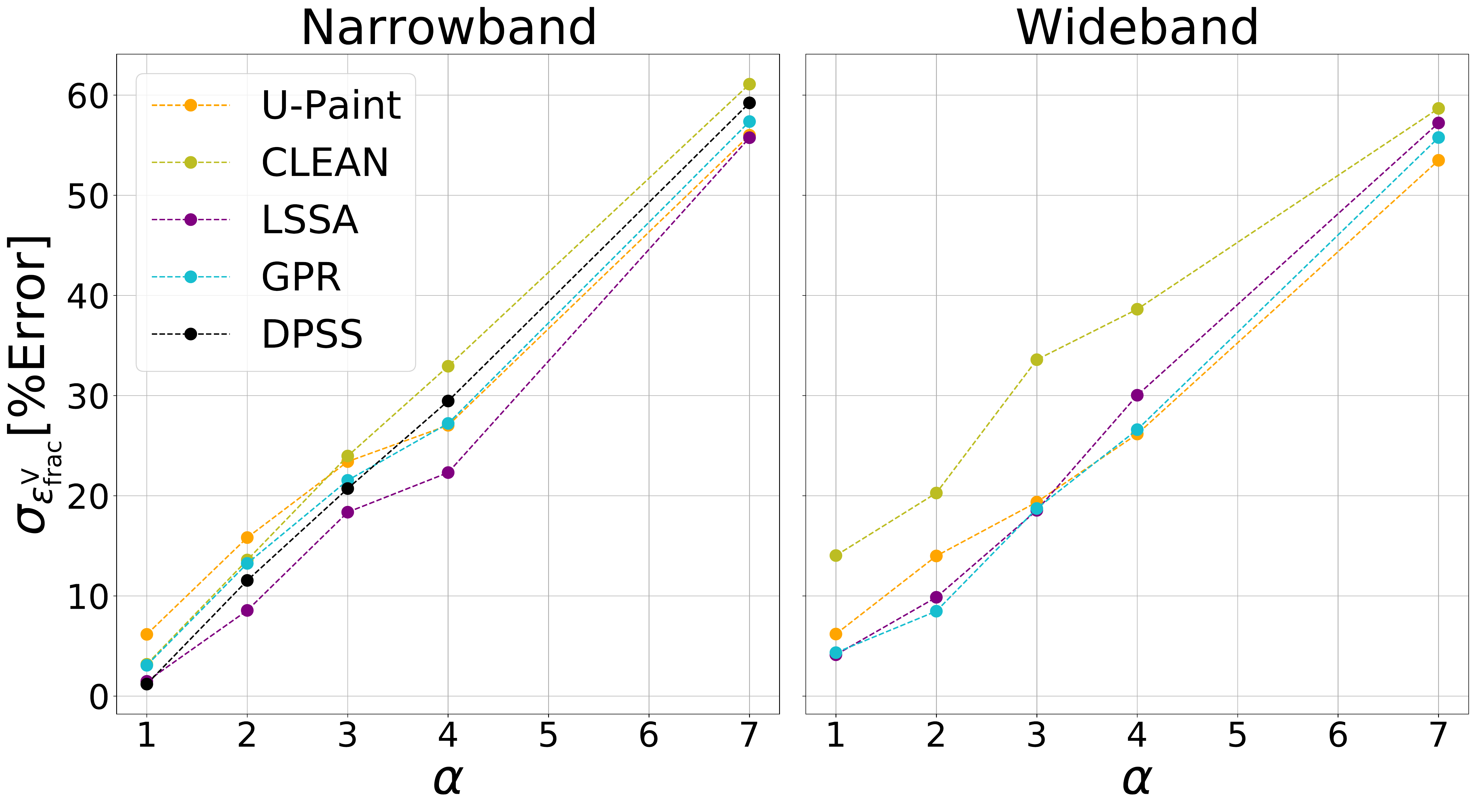}
  \caption{The standard deviation of the fractional error in the visibilities $\epsilon^{\rm V}_{\rm frac}$ as a function of the thermal noise level in the visibilities. The parameter $\alpha$ is used as a proxy for the thermal noise level (see Equation \ref{eq:Radiometer}).} 
  \label{fig:Sec5pt4PerformanceSigma}
\end{figure}


 Since the inpainting techniques can not predict exact noise realisations in the dataset, we expect an increase in the amplitude of the fractional errors. In Figure \ref{fig:Sec5pt4PerformanceSigma} we show the evolution of the standard deviation of the fractional error (in percentage of the true visibilities) in the wideband and narrowband regions of the visibilities as a function of thermal noise level in the visibilities. We use the dimensionless parameter $\alpha$ as a proxy for the thermal noise level in the dataset (see Equation \ref{eq:Radiometer}). Notice the linear evolution of $\sigma_{\epsilon^{\rm V}_{\rm frac}}$ with $\alpha$. This shows that the standard deviation of the fractional error is linearly proportional to the standard deviation of the noise level in the dataset. Thus as one averages down $\alpha$ through LST binning (or equivalently, other types of averaging), the performance of the inpainting techniques improves linearly. Therefore performing the inpainting before the LST binning in a data analysis pipeline will result in the same performance. In contrast, a non-linear evolution of $\sigma_{\epsilon^{\rm V}_{\rm frac}}$ with $\alpha$ would describe a scenario where the $\sigma_{\epsilon^{\rm V}_{\rm frac}}$ depends on the standard deviation of the noise beyond just simple sample variance of the noise, i.e. there may be advantages to applying the inpainting technique at a specific noise level before or after LST binning (depending on whether the relationship between $\sigma_{\epsilon^{\rm V}_{\rm frac}} $ and $\alpha$ is more or less steeper than linear). Thus Figure \ref{fig:Sec5pt4PerformanceSigma} reinforces our assertion that each inpainting technique captures only the underlying sky signal of the dataset. 

Building on the intuition of the error properties in the visibilities, we now examine errors in the power spectrum derived from the inpainted visibilities and form connections between the errors of both components.

\section{Power Spectrum Error Characterization}
\label{sec:PowerSpectrumErrorCharacterization}
 In this section we characterize the type of errors in $P(\tau)$ due to the inpainting as well as establish the relationship between the errors in the model visibilities and their corresponding delay power spectra. We propagate two versions of the visibilities through the power spectrum. The true visibilities (which do not have any corrupted regions), and the corrupt visibilities (where inpainted models have been replaced in the RFI corrupt regions). Thus we have the power spectrum derived from the model visibilities $P_{\rm model}$, and the true power spectrum  $P_{\rm true}$ derived from the true visibilities. We can define the residuals analogously to Equations \ref{eq:frac_errors_def} , i.e $\epsilon^{\rm P}_{\rm r} = P_{\rm model} - P_{\rm true}$. Similarly for the fractional errors $\epsilon^{\rm P}_{\rm frac} = (P_{\rm model} - P_{\rm true})/P_{\rm true}$. We separate our analysis in terms of delay modes ($\tau$) inside and outside the wedge. 
 This section is structured as follows. In Section \ref{ssec:RealisticSpecWinSec64} we discuss the properties of the power spectra derived from the model visibilities.
 In Section \ref{sssec:relationship_between_pspec_vis_errs} we establish a relationship between the errors in the model visibilities from Section \ref{sec:BigSectionErrorQuantification} and the model power spectra from Section \ref{ssec:RealisticSpecWinSec64}. 

\subsection{P1V Spectral Window}
\label{ssec:RealisticSpecWinSec64}
 \begin{figure}
  \includegraphics[width=0.5\textwidth]{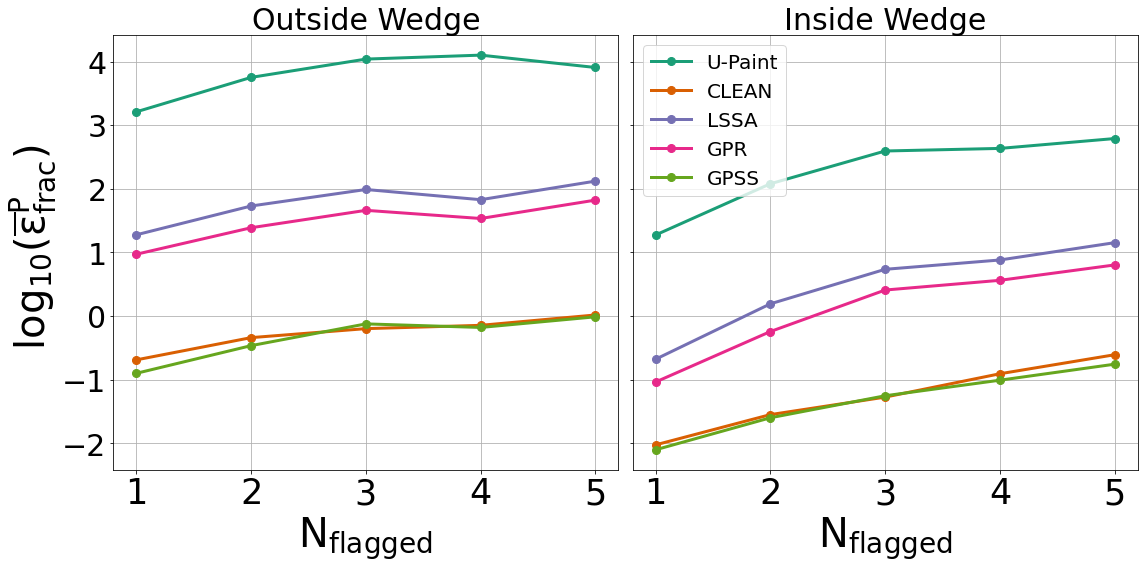}
  \caption{The fractional errors in the wedge modes (left) and non-wedge modes (right) of inpaint model power spectra $\epsilon^{\rm P}_{\rm frac}$ as a function of the number of flagged channels within the spectral window. The P1V spectral window is used to estimate the power spectra. }
  \label{fig:Section6pt4Nflagged}
\end{figure}

\begin{figure*}
  \includegraphics[width=0.8\textwidth]{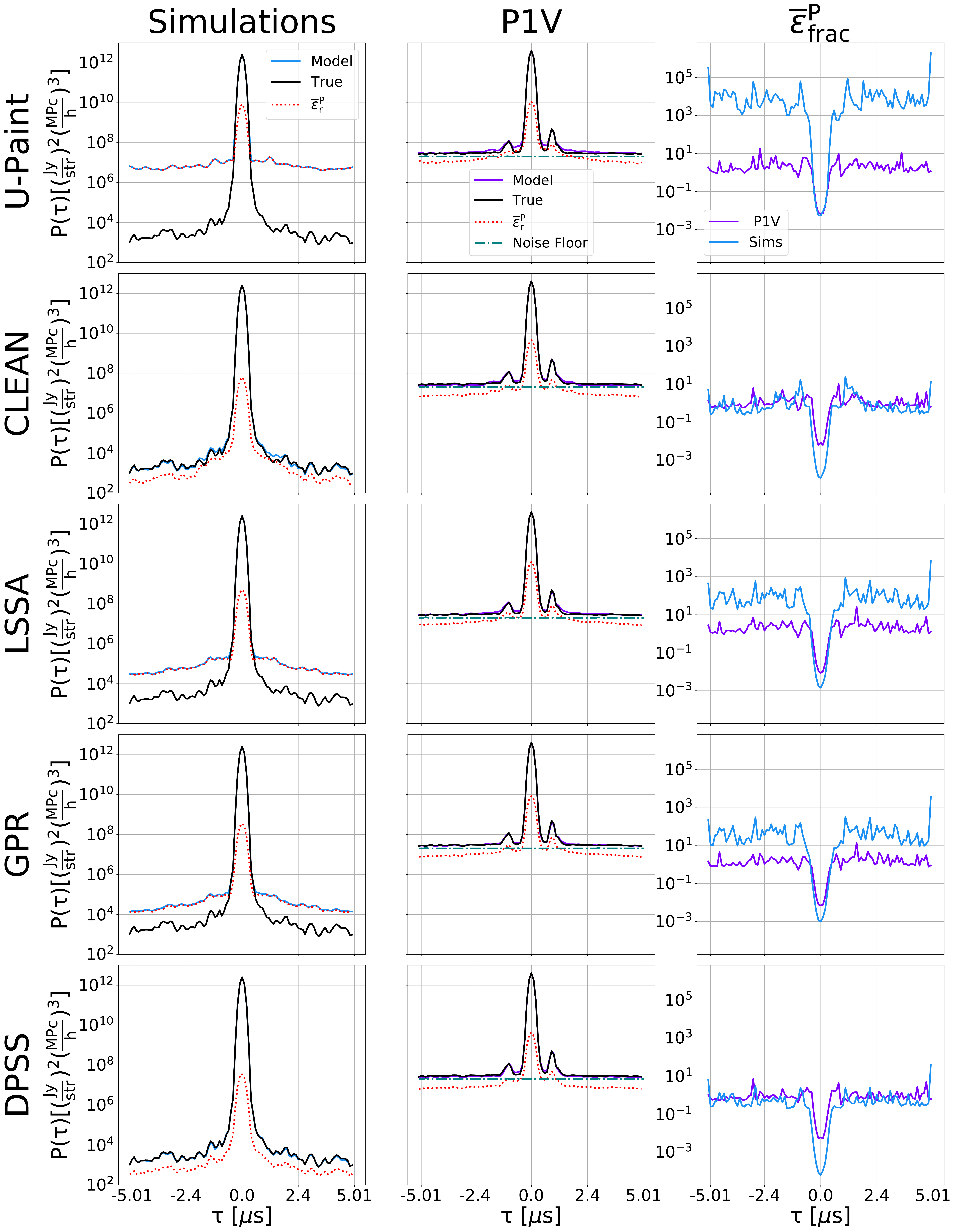}
  \caption{Each inpainting technique is applied to the simulated data discussed in Section \ref{sssec:P1V_Data}. The P1V spectral window is used to estimate the inpaint model power spectra. Left column: blue curves correspond to inpaint model power spectra. The black curves correspond to the true power spectra and the red dotted curves correspond to the residuals. Each row corresponds to a different inpaint technique used to inpaint RFI flagged simulated visibilities. Second column (see Section \ref{ssec:P1VResultsPspec}): Same as first column but with real P1V data. Purple curves correspond to inpaint model power spectra and black curves the true power spectra. Red curves are the residuals. The third column corresponds to the fractional errors $\epsilon^{\rm P}_{\rm frac}$ in inpaint model power spectra from simulated data (blue) and the P1V data (purple). The dotted teal line corresponds to the power spectrum of the thermal noise floor of P1V data \citep{HERAH1CData}. }
  \label{fig:Sec7p4PspecModelsWindowII}
\end{figure*}

 \begin{figure*}
  \includegraphics[width=0.85\textwidth]{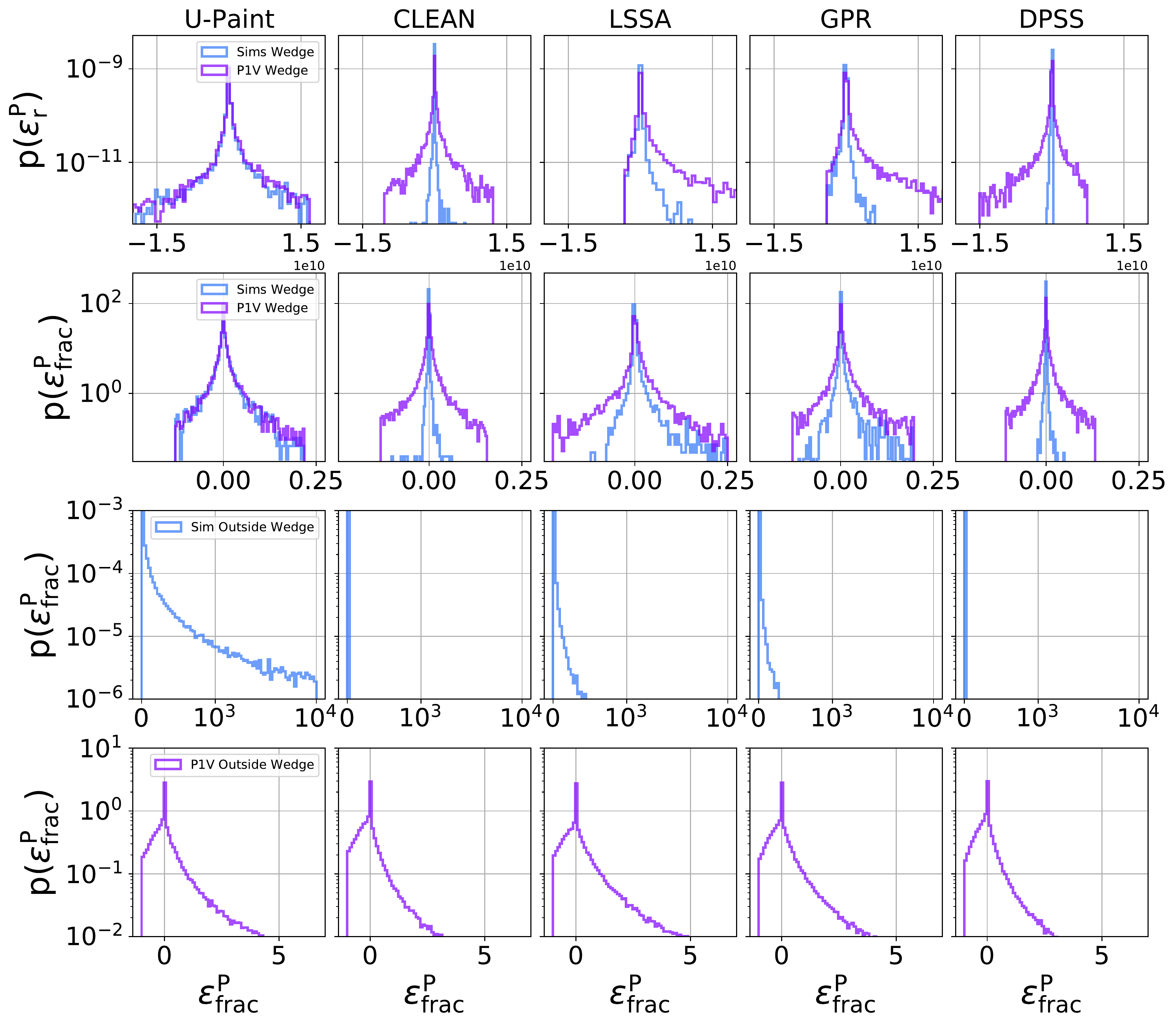}
  \caption{Blue curves correspond to simulated data and purple curves correspond to P1V data. Distribution of residuals (first row) and fractional errors (second through fourth rows) in the inpaint model power spectra. Residuals are shown for wedge modes while the fractional errors are separated according to $\tau$ modes lying inside the wedge (second row) and outside the wedge (third and fourth rows). Third row corresponds to simulated data while the fourth row corresponds to P1V data.  }
  \label{fig:PspecDistributionsWindowII}
\end{figure*}

 \begin{figure*}
  \includegraphics[width=0.95\textwidth]{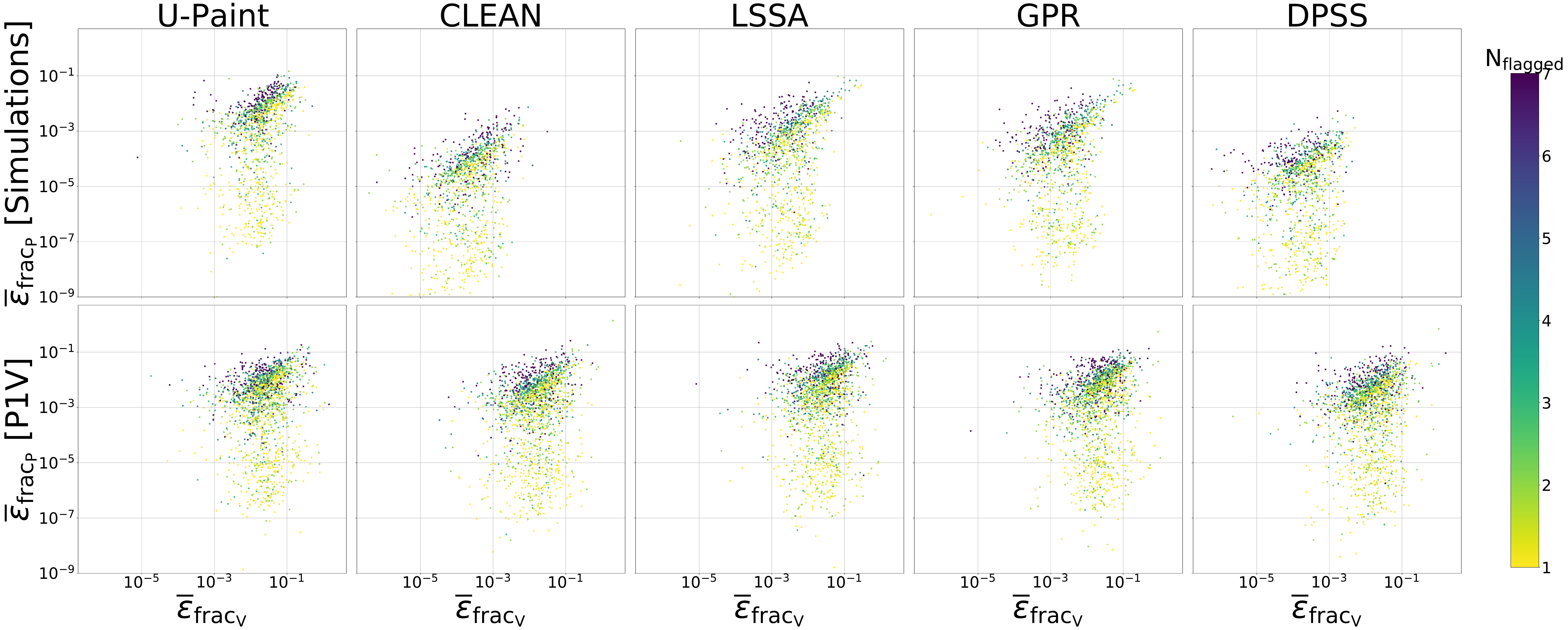}
  \caption{Relationship between the mean fractional errors in the inpaint model visibilities $\overline{\epsilon}^{\rm V}_{\rm frac}$ and the mean fractional errors in their corresponding power spectra $\overline{\epsilon}^{\rm P}_{\rm frac}$. We compute the mean fractional error of the inpaint models in RFI flagged frequency channels within the P1V spectral window. This process is repeated at each LST. Their corresponding power spectra are estimated using the same P1V spectral window. The mean of the fractional errors in the model power spectra are computed using $\tau$ modes inside the wedge. Each LST is plotted as a scatter point. The LSTs are colour coded according to the number of flagged frequency channels at that LST. In the top row this procedure is applied to simulated data while in the bottom row this procedure is applied to P1V data. }
  \label{fig:ScatterNflaggedWindowIIP1VSims}
\end{figure*}

We compute the power spectra using the spectral window from $119$MHz to $129$MHz which is one of the spectral windows used to set upper limits on the power spectrum in HERA Phase 1 Upper Limits. This window contains both flagged and non-flagged regions of the visibilities. Recall that in our example HERA flags in Figure \ref{fig:HERAFlags}, this frequency range spans over 100 channels and corresponds to a region of the visibilities with only narrowband RFI. In this spectral range, the number of flagged channels at each LST range from 0-31 frequency channels which  corresponds to up to $31\%$ of the spectral window used to compute the power spectra. Recall that the power spectrum is computed independently at each LST and thus there are LSTs where one third of the band is flagged and LSTs without any flags at all. We restrict our analysis to LSTs with at least one flagged channel. This reduces the number of sample power spectra with which we can form our analysis. We find that the key indicator of performance are the number of flagged pixels within the band. Denote the number of flagged channels at each LST by $N_{\rm flagged}$. When $N_{\rm flagged} = 0$ we have no errors in $P(\tau)$. As we increase $N_{\rm flagged}$ a larger fraction of the spectral window are flagged. For fixed $N_{\rm flagged}$, the arrangement of the RFI also affects the performance. For example, scenarios with four consecutively flagged channels does not yield similar errors as when the four flagged channels are dispersed. Denote $N_{\rm max_{\rm_c}}$ as the number of consecutively flagged channels. When $N_{\rm max_{\rm_c}}$ increases we eventually have a wideband feature which have greater fractional errors relative to narrowband RFI. Thus power spectrum estimates derived from wideband RFI features in the visibilities have drastically increased errors relative power spectrum estimates derived from regions of the visibilities with intermittent (i.e narrowband) RFI. 
Thus both $N_{\rm flagged}$ and their arrangement within the spectral window will affect the errors in the model power spectra. For this analysis we examine the effect of $N_{\rm flagged}$ on the model power spectra, i.e. we treat $N_{\rm flagged}$ as the dominant effect and $N_{\rm max_{\rm_c}}$ as a secondary effect which we leave to future work. In Figure \ref{fig:Section6pt4Nflagged} we show the mean fractional errors of the model power spectrum $\overline{\epsilon}^{\rm P}_{\rm frac}$ as a function of $N_{\rm flagged}$ separated by modes outside and inside the wedge. Note that the smallest mean fractional error $\overline{\epsilon}^{\rm P}_{\rm r}$ occurs when only one pixel is flagged. In our flags, $25\%$ of all LSTs have only one flagged channel. The mean fractional errors in both wedge and non-wedge modes of the model power spectra increase rapidly as a function of the number of flagged regions for $N_{\rm flagged} < 5$. By $N_{\rm flagged} = 5$ the fractional errors for modes outside and inside the wedge are an order of magnitude greater than when only one channel is flagged. On average, $90\%$ of the LSTs in HERA flags have 5 flagged channels or less. Thus most LSTs fall within this error range. 


We now look at the model power spectra after averaging over LST. This implicitly averages over $N_{\rm flagged}$. We ignore LSTs that have don't have any flagged channels. In the first column of Figure \ref{fig:Sec7p4PspecModelsWindowII} we LST average the model power spectrum (blue curve) and compare it to the LST averaged true power spectrum (black curve). The dotted red curve corresponds on the mean residuals $\overline{\epsilon}^{\rm P}_{\rm r}$ between the model power spectra and the true power spectra. 
Referring to the fractional errors in the blue curves of the third column, we can see that CLEAN and DPSS produce power spectra models with the smallest fractional errors in the wedge, followed by GPR, LSSA and U-Paint. By examining the larger errors in $P_{\rm model}$ for the largest delay modes, it is clear that none of inpainting methods inpaint noise. We can see that the inpainting techniques only capture the sky signal. This leads to larger errors in the largest $\tau$ modes which are noise dominated. CLEAN and DPSS models have fractional errors on the order $\sim 10^0$, while GPR and LSSA are on the order $10$ and U-Paint on the order $10^4$. This is the due to the fine frequency structure imprinted into the visibilities by U-Paint (see Section \ref{sec:BigSectionErrorQuantification}). Note that analysis of the errors the largest $\tau$ modes of $P_{\rm model}$ are only possible since we are using simulated data, which is systematic free, and less noisy than real data. In the future we will continue to make progress on reducing systematics in our data, thus increasing the importance of understanding the behaviour of inpaint models in the largest $\tau$ modes. In that scenario, spectral structure imprinted into model power spectra by inpaint methods such as U-Paint must be accounted for.

In the top row of Figure \ref{fig:PspecDistributionsWindowII} we show the distribution of residuals errors $p(\epsilon^{\rm P}_{\rm r})$ for modes inside the wedge (blue solid curves). The residuals are smallest for DPSS and CLEAN inpainting techniques. In the second row of Figure \ref{fig:PspecDistributionsWindowII} we show the distribution of fractional errors $p(\epsilon^{\rm P}_{\rm frac})$ for wedge modes (solid blue curves) where we again see that DPSS and CLEAN have the smallest range of fractional errors. We find that the profile of $p(\epsilon^{\rm P}_{\rm frac})$ for modes inside the wedge are best described by a generalized normal distribution. For modes outside the wedge (third row in Figure \ref{fig:PspecDistributionsWindowII}), LSSA, U-Paint and GPR are characterized by a log normal distribution. Recall that for the $\tau$ modes outside the wedge, $P_{\rm model} \gg P_{\rm true}$ for U-Paint, LSSA and GPR. Thus their fractional error distributions are composed of samples with $\epsilon^{\rm P}_{\rm frac} \gg 0$. This gives the distribution long positive tails. \footnote{Lognormal distributions are only defined for positive values and have long tails making this profile ideal to describe the non-wedge modes of these inpainting techniques}. Since CLEAN and DPSS have much smaller errors outside the wedge, their distributions $p(\epsilon^{\rm P}_{\rm frac})$ are confined to $p(\epsilon^{\rm P}_{\rm frac}) < 10$. 

\subsection{Relationship Between Visibility and Power Spectrum Errors}
\label{sssec:relationship_between_pspec_vis_errs}

In Sections \ref{sssec:CLEANandLSSA_VisibilityAnalysis} and \ref{ssec:RealisticSpecWinSec64} we discussed the error characteristics of the model visibilites and model power spectra. Since the model power spectra are derived from the model visibilities, we expect a relationship to exist between their errors. Since the errors in $P_{\rm model}(\tau)$ are different for modes inside and outside the wedge, we expect the relationship between model visibilities and model power spectra to also depend on $\tau$. In this Section we explore these relationships. 

Consider the 100 frequency channels spanning the frequencies $119$MHz-$129$MHz corresponding to our spectral window. A direct relationship between the errors in each pixel of the model visibilities and the corresponding model power spectra is impractical since the power spectrum is derived from all frequency channels within this spectral window. We therefore find it convenient to establish a relationship between the mean power spectrum errors and the mean amplitude errors of the visibilities. Since the inpaint models do not inpaint noise, and since the large $\tau$ modes are noise dominated, we establish a relationship between the mean fractional errors of the visibilities $\overline{\epsilon}^{\rm V}_{\rm frac}$, and the mean fractional errors in the wedge modes of their corresponding power spectra $\overline{\epsilon}^{\rm P}_{\rm frac}$. The mean fractional error in the visibilities are given by
\begin{equation}
    \label{eq:SampleAndAverageVisibilitiesFrac}
    \overline{\epsilon}^{\rm V}_{\rm r}(\rm LST)  = \frac{1}{N_{\rm flagged}} \sum^{i = 129}_{i = 119} \left [ \frac{V_{\rm model}(\rm LST, \nu_i) - V_{\rm true}(\rm LST, \nu_i) }{V_{\rm true}(\rm LST, \nu_i)} \right] .
\end{equation}
The averaging in Equation \ref{eq:SampleAndAverageVisibilitiesFrac} occurs along the frequency domain which leaves us with $N_{\rm LST}$ samples. This translates to $\sim 5000$ samples in our simulation data. The mean fractional error for the model power spectrum are similarly computed:
\begin{equation}
    \label{eq:SampleAndAveragePspecFrac}
    \overline{\epsilon}^{\rm P}_{\rm r}(\rm LST)  = \frac{1}{7} \sum^{i = \tau_g}_{i = -\tau_g} \left [ \frac{P_{\rm model}(\rm LST, \tau_i) - P_{\rm true}(\rm LST, \tau_i) }{P_{\rm true}(\rm LST, \tau_i)} \right]
\end{equation}
where the index $i$ tracks the $\tau$ bins in the wedge modes of the power spectrum and $7$ corresponds to the number of $\tau$ modes inside the wedge. The averaging in Equation \ref{eq:SampleAndAveragePspecFrac} occurs along the $\tau$ domain which leaves us with $N_{\rm LST}$ samples. 
For intuition we can explore an analytical relationship between $\overline{\epsilon}^{\rm V}_{\rm frac}$ and $\overline{\epsilon}^{\rm P}_{\rm frac}$. If we approximate the wedge modes of Equation \ref{eq:SampleAndAveragePspecFrac} as being uniform and equal to the error in $P(\tau = 0)$ then we can approximate Equation \ref{eq:SampleAndAveragePspecFrac} as
\begin{equation}
    \label{eq:pspecvisibrelationshipmotivation}
    \overline{\epsilon}^{\rm P}_{\rm frac} = \left ( \frac{\overline{P}_{\rm model} - \overline{P}_{\rm true}}{\overline{P}_{\rm true}} \right )_{\tau = 0} = \frac{\overline{V}^2_{\rm model}-\overline{V}^2_{\rm true} }{\overline{V}^2_{\rm true} } .
\end{equation}
where the last step is due to $P(\tau = 0)$ corresponding to the square mean of the visibilities. Therefore we can rewrite the right side of Equation \ref{eq:pspecvisibrelationshipmotivation} as 
\begin{equation}
    \label{eq:pspecvisibrelationshipmotivation2}
     \overline{\epsilon}^{\rm P}_{\rm frac} =  \overline{\epsilon}^{\rm V}_{\rm frac}\left( \frac{ \overline{V}_{\rm model} + \overline{V}_{\rm true} }{\overline{V}_{\rm true}} \right ) .
\end{equation}
In scenarios where the mean of the model visibilities $\overline{V}_{\rm model}$ is consistently related to the mean of the true visibilities $\overline{V}_{\rm true}$ by a constant $\delta$, we can write $\overline{V}_{\rm model} = \delta \overline{V}_{\rm true}$. This is not a bad assumption for LSTs where the amplitude of the visibilities are relatively constant. For example in Figure \ref{fig:5pt2AmplitudePredictionsAll} we can see that the fractional error remains reasonably uniform in LSTs in the vicinity of 119MHz to 129MHz. In this situation Equation \ref{eq:6pt3ForegroundPowerLaw} can be recast as 
\begin{equation}
    \label{eq:6pt3ForegroundPowerLaw}
     \overline{\epsilon}^{\rm P}_{\rm frac} = (1 + \delta) \overline{\epsilon}^{\rm V}_{\rm frac} ,
\end{equation}
which suggests the mean fractional error in the power spectrum $\overline{\epsilon}^{\rm P}_{\rm frac}$ scale linearly with the mean fractional error in the amplitude of the visibilities. Note that we expect this approximation to no longer be valid as the largest $\tau$ modes are included into the mean fractional errors of Equation \ref{eq:SampleAndAveragePspecFrac}. In the top row of Figure \ref{fig:ScatterNflaggedWindowIIP1VSims} we show the relationship between $\overline{\epsilon}^{\rm P}_{\rm frac}$ and $\overline{\epsilon}^{\rm V}_{\rm frac}$ where each scatter point corresponds to an individual LST. From the previous section, we also expect that the relationship between $\overline{\epsilon}^{\rm V}_{\rm frac}$ and $\overline{\epsilon}^{\rm P}_{\rm frac}$ will depend on the number of flagged channels at each LST. We colour code the scatter points according to the number of flagged channels at that LST. Note that LSTs with $N_{\rm flagged} = 1$ (the brightest green, and smallest points in Figure \ref{fig:ScatterNflaggedWindowIIP1VSims}) are located at the smallest values of $\overline{\epsilon}^{\rm P}_{\rm frac}$ indicating that these LSTs produce the smallest mean errors in $P(\tau)$. It is also clear that LSTs with $N_{\rm flagged} = 1$ don't appear to strongly cluster together, or follow the same cohesive relationship as when $N_{\rm flagged} > 1$ . This is likely due to sample variance, since the mean fractional errors in the visibility and power spectrum are computed using a single channel making $\overline{\epsilon}^{\rm P}_{\rm frac}$ and $\overline{\epsilon}^{\rm V}_{\rm frac}$ prone to scatter. Conversely, LSTs with $N_{\rm flagged} \gg 1$ appear to follow a clearer linear trend. We can also see that LSTs with $N_{\rm flagged} > 20$ tend to produce the largest values of $\overline{\epsilon}^{\rm P}_{\rm frac}$. 

\section{Application to Phase 1 HERA Data}
\label{sec:RealDataBigSection}
In Sections \ref{sec:BigSectionErrorQuantification} \& \ref{sec:PowerSpectrumErrorCharacterization} we discussed the performance of each inpainting technique as well as the types of errors they introduce as part of computation of the power spectrum. However the analysis was performed on simulated data. While our  simulated data from Section \ref{sssec:P1V_Data} does take into account the instrument, it doesn't fully capture all the instrumental effects such as systematics that come along with a real observation. In this section we characterize the errors introduced in an actual HERA analysis pipeline. To do this we apply U-Paint, CLEAN, LSSA, GPR, and DPSS to the P1V HERA data discussed in Section \ref{sssec:P1V_Data} and repeat our analysis from Sections \ref{sec:BigSectionErrorQuantification} and \ref{sec:PowerSpectrumErrorCharacterization}. To keep our analysis as similar as possible to the true HERA analysis pipeline, we use the 119MHz-129MHz spectral window to compute the power spectra. In order to quantify the errors in $V_{\rm model}$ and $P_{\rm model}$ using the same methods in the previous sections, the true (i.e. known) visibilities and power spectrum are required. One hurdle in realizing this goal is that since the true solution to the RFI flagged regions of real P1V data do not exist, therefore we need to modify our analysis procedure. In Section \ref{sss:P1VConfig} we discuss our modifications to the procedure outlined in Section \ref{sec:BigSectionErrorQuantification}. In Sections \ref{ssec:P1VResultsV} and \ref{ssec:P1VResultsPspec} we discuss our results showing that our intuition and error characterization carries over from the previous sections and thus we can infer the error properties in the true analysis from simulation.

\subsection{Flagged Regions \& Analysis Configuration}
\label{sss:P1VConfig}

 \begin{figure*}
  \includegraphics[width=0.95\textwidth]{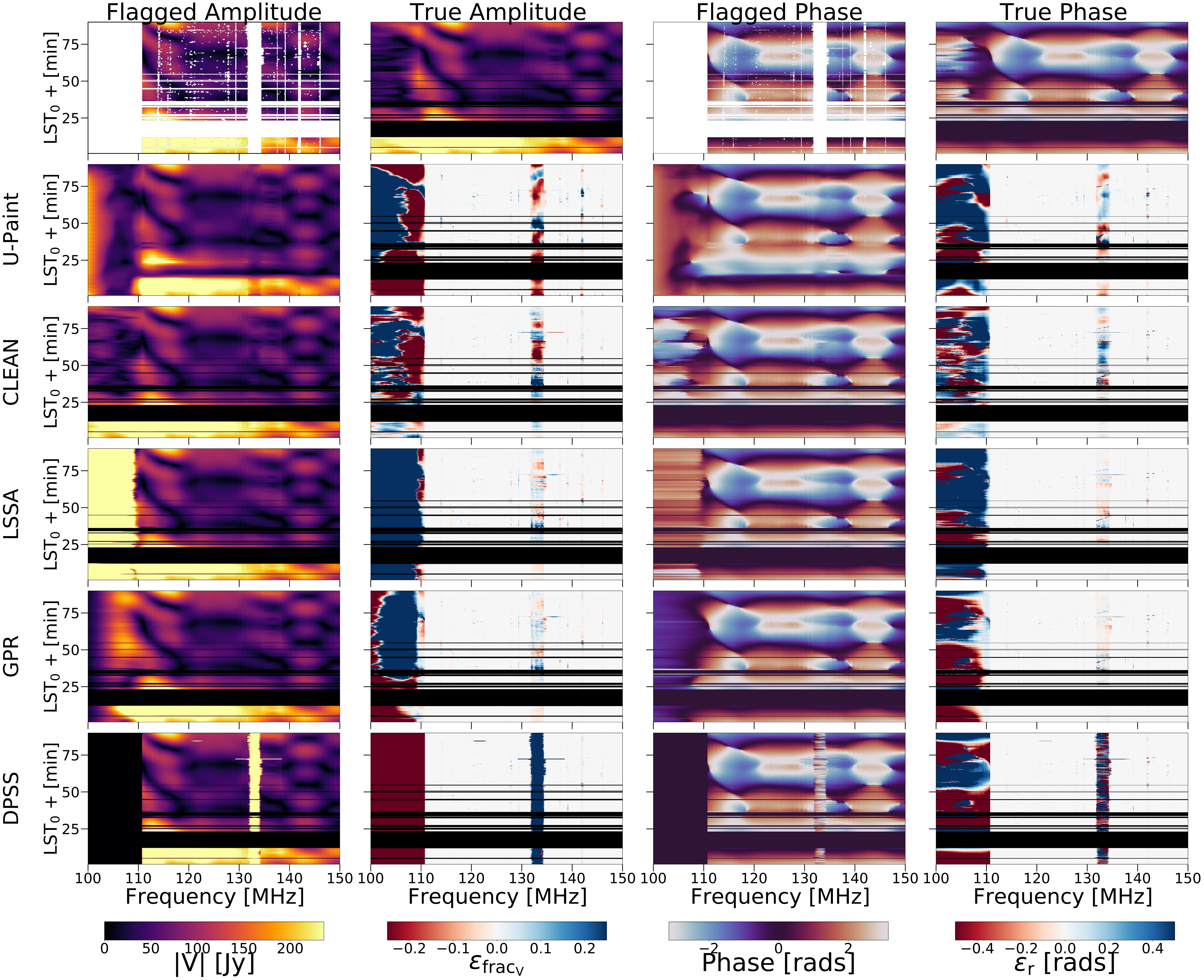}
  \caption{Same as Figure \ref{fig:5pt2AmplitudePredictionsAll} but with the P1V visibilities from Section \ref{sssec:P1V_Data}. The true visibilities in the first row (second and fourth column) have been initially inpainted with the CLEAN algorithm to generate placeholder data for the RFI flagged regions. The inpaint techniques are then applied to a set of flags which are shifted $40$ channels to the left. This is done in order to avoid inpainting over the already CLEANed data. See Section \ref{sec:RealDataBigSection} for more details regarding our procedure. Note that as compared to Figure \ref{fig:5pt2AmplitudePredictionsAll} the fractional errors in the model visibilities increase. 
}
  \label{fig:sec8pt2PspecModels}
\end{figure*}

 Denote the flagged regions of the P1V visibilities as $M_{\rm P1V}$. To apply the error metrics discussed in Section \ref{sssec:CLEANandLSSA_VisibilityAnalysis}, the ``true'' visibilities in $M_{\rm P1V}$ are required to be known. This is not the case for $M_{\rm P1V}$ regions of P1V data.  
 This causes several difficulties and prevents us from directly repeating our analysis procedure from Sections \ref{sec:BigSectionErrorQuantification} and \ref{sec:PowerSpectrumErrorCharacterization}. Furthermore, the presence of RFI flags can introduce artifacts into the power spectrum due to the Fourier transforming the sharp discontinuities between flagged and unflagged regions. To avoid introducing these artifacts into the inpaint models of U-Paint, CLEAN, LSSR, GPR and DPSS we inpaint over the flagged regions of the P1V data using the CLEAN algorithm. We use the CLEAN parameter values that were used in \cite{HERAH1CData}. After this step the flagged regions have been replaced with CLEAN inpaint models. Repeating our error analysis on the $M_{\rm P1V}$ flagged regions of P1V data now means that we would be using the CLEAN models as the true visibilities (which we wish to avoid). We therefore create a new set of flags by taking $M_{\rm P1V}$ and shifting them over in frequency space by 40 channels. We refer to the shifted flags which are applied to the visibilities as $M_{\rm shift}$.  Applying our analysis on using $M_{\rm shift}$ rather than $M_{\rm P1V}$ allows us to use regions of the visibilities where the true values are known as well as to keep the structure of the real P1V flags. This procedure is not perfect in that there is an overlap of some of the narrowband RFI in the $M_{\rm shift}$ and $M_{\rm P1V}$. However $< 5\%$ of the narrowband RFI in $M_{\rm shift}$ overlaps with narrowband RFI in $M_{\rm P1V}$. This estimate does not include the wideband features below $110$MHz and above $174$MHz. In such overlapping channels, the true solution is therefore CLEAN inpaint model. Since the overlap percentage is small, we don't expect this overlap to significantly influence our results. Note that by applying this shifting procedure, certain characteristic broadband RFI features of $M_{\rm shift}$ no longer align with their corresponding frequency bins. For example the ORBCOMM feature is characteristically found at $136$MHz. Conversely, narrowband RFI is intermittent, and thus $M_{\rm shift}$ flags provides us with a statistically representative set of narrowband RFI samples.

To generate the inpainted models for the flagged regions, i.e $M_{\rm shift}$ using U-Paint, we consider two network configurations. Each scenario produces comparable results. In the first case we use the weights of the network which has been trained on the simulated data described in Section \ref{sssec:P1V_Data} (at the fiducial noise level). This is the network which was used in the analysis throughout Sections \ref{sec:BigSectionErrorQuantification} and \ref{sec:PowerSpectrumErrorCharacterization}. For completeness, and to examine the range performance that can be obtained by our network, we try a second scenario. In the second scenario we retrain the network on P1V data after having performed the CLEAN procedure described above. Thus in this scenario an initial CLEAN is still performed and $M_{\rm shift}$ are used as our flagged regions. We find that both scenarios produce comparable results on the P1V data. We thus use the network from scenario 1 (i.e. the network which was used in the analysis throughout Sections \ref{sec:BigSectionErrorQuantification} and \ref{sec:PowerSpectrumErrorCharacterization} ) to generate inpaint models. To generate inpaint models for CLEAN, LSSA, GPR, and DPSS we use the same parameters described in Section \ref{sec:InpaintingTechniques} for the simulated data at the fiducial noise level. 

\subsection{Results}
\label{ssec:P1VResultsV}

\begin{table}
\caption{Summary of key error metrics for the amplitude component of P1V visibilities.  } 
\label{tab:P1VDataVis}
\begin{center}
\begin{tabular}{|c|c|c|c|c|} 
\hline

Error & $\overline{\sigma}_{{\epsilon}_{\rm frac}}$ & $\overline{\sigma}_{{\epsilon}_{\rm frac}}$  & $\overline{\mu}_{{\epsilon}_{\rm frac}}$ & $\overline{\mu}_{{\epsilon}_{\rm frac}}$\\ 
  RFI       &   Narrowband   &   All   & Narrowband & All          \\
\hline\hline
U-Paint & 24.5\% & 98.7\% &  2.1\% &  4.9\%\\
\hline
CLEAN & 19.1\% & 58.2\% &  0.81\% &  5.4\% \\ 
\hline
LSSA & 44\% & 81.2\% &  1.9\% &  3.6\%\\ 
\hline
GPR & 19.2\% & 41.3\% &  0.65\% &  2.1\%\\ 
\hline
DPSS & 15\% & -&  0.5\% &  -\\ 
\end{tabular}
\end{center}
\end{table}

 \begin{figure*}
  \includegraphics[width=0.95\textwidth]{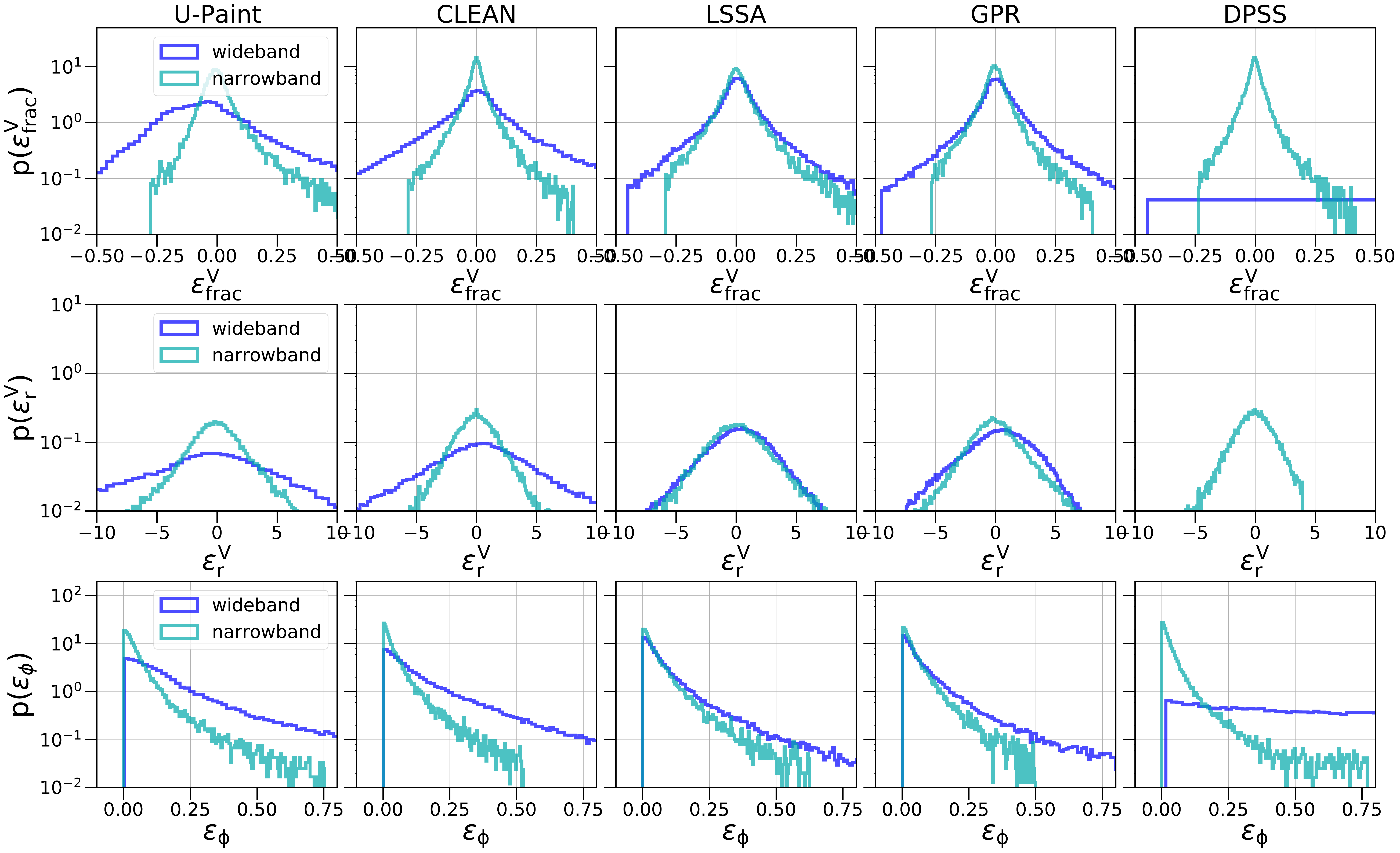}
  \caption{First row: distribution of fractional errors in the amplitude of P1V visibilities. Second row: distribution of residuals in P1V visibilities. Third row: distributions of phase errors $\epsilon_\phi$ in the phase of P1V visibilities. In each case the blue curves correspond to distributions constructed using wideband RFI samples only. Teal curves correspond to distributions constructed using narrowband RFI samples only}.
  \label{fig:sec8pt2VisibHistograms}
\end{figure*}

In Figure \ref{fig:sec8pt2PspecModels} we show an example image of RFI flagged P1V data which has been inpainted. The first panel in the first row corresponds to the P1V visibilities with $M_{\rm shift}$ applied. The first panel in the second row corresponds to the P1V visibilities after an initial CLEAN inpaint, from here onward we refer to this as the ``true'' visibilities. Note that the LSTs where all frequency channels are flagged have unknown true visibilities and haven't been inpainted over since CLEAN avoids these LSTs. Therefore the ``true'' visibilities in the upper left panel of Figure \ref{fig:sec8pt2PspecModels} still appear to have flagged regions. The visibilities where RFI flags have been reapplied are on the upper right. Each subsequent row corresponds to the indicated inpainted model (left) and their fractional errors (right). Note that U-Paint still inpaints over LSTs with no data, however since a fractional error cannot be computed (true visibilities are unknown), we do not display a fractional error. In the two last columns of Figure \ref{fig:sec8pt2PspecModels} we show the corresponding phase component of the visibilities. Referring to the fractional errors of the amplitude components and residuals in the phase component of Figure \ref{fig:sec8pt2PspecModels} we can see that the inpainting methods again perform better in the narrowband regions as compared to the wideband regions. Notice that the residuals in the phase component are much larger than their simulated counterparts in Figure \ref{fig:5pt2AmplitudePredictionsAll}. Similarly comparing the fractional errors in second column of Figure \ref{fig:sec8pt2PspecModels} to the fractional errors of the inpainted model of the simulated data in \ref{fig:5pt2AmplitudePredictionsAll}, we see that there are larger fluctuations in fractional error in the P1V inpainted models relative to the simulated data. This is the case for each inpaint method. The standard deviations and the mean of the fractional errors are summarized in Table \ref{tab:P1VDataVis}. 

In Figure \ref{fig:sec8pt2VisibHistograms} we show the probability density function of the fractional errors $p(\epsilon^{\rm V}_{\rm frac})$ (top row), residuals $p(\epsilon^{\rm V}_{\rm r} )$ (middle row) and the distribution of errors $p(\epsilon_\phi)$ for the phase component of the visibilities (bottom row) as a function of the type of flags, i.e. narrowband and wideband. Focusing on the top row, we can see that the profile of the probability distributions functions $p(\epsilon^{\rm V}_{\rm frac})$ share many qualitative characteristics with their corresponding distributions from Section \ref{sssec:CLEANandLSSA_VisibilityAnalysis}. For example, we can again see that DPSS still produces the most accurate results for narrowband RFI followed by CLEAN, GPR, LSSA and U-Paint. However by comparing the extent of the distributions for narrowband RFI, we can see the performances are less discrepant. By examining the range of errors we can see that GPR and LSSA produce the smallest range of fractional errors for narrowband RFI.

In the second row of Figure \ref{fig:sec8pt2VisibHistograms} we show the  distribution of residuals $p(\epsilon^{\rm V}_{\rm r})$ for each inpainting technique. Through comparison with the middle row in Figure \ref{fig:5pt3HistogramMegaPlot} we can see that the residuals using the P1V data are larger than those using the simulated data. As was the case with the distribution of fractional errors $p(\epsilon^{\rm V}_{\rm frac})$ from above, we can see that the maximum range of residuals in narrowband RFI are similar among the inpainting techniques. 
For each inpainting technique, we find that the profile of $p(\epsilon^{\rm V}_{\rm r})$ and $p(\epsilon^{\rm V}_{\rm frac})$ are best characterized by a generalized normal distribution.

In the bottom row of Figure \ref{fig:sec8pt2VisibHistograms} we show the distribution of errors $\epsilon_\phi$ in the phase component of the P1V inpaint models. We can see that relative to the distributions $p(\epsilon_\phi)$ in Figure \ref{fig:5pt3HistogramMegaPlot} which were generated with simulated data there is an apparent performance decrease when applying the inpainting techniques to P1V data. For narrowband RFI, we find that the tails extend into the range $\epsilon_\phi \sim 0.75$rads while the tails of $p(\epsilon_\phi)$ in wideband RFI regions extend into the range $\epsilon > \pi/3$ which reflects a more significant deviation in phase relative to the true values.  Unlike the distributions in Figure \ref{fig:5pt3HistogramMegaPlot} which were generated with simulated data, U-Paint does show consistent performance in the phase component.  Similar to Section \ref{sssec:CLEANandLSSA_VisibilityAnalysis}, we find that all distributions functions are best described by a log normal distribution.

\subsection{Power Spectrum}
\label{ssec:P1VResultsPspec}

In this section we compute the power spectrum of the inpaint models. To do so we use the P1V spectral window. In the middle column of Figure \ref{fig:Sec7p4PspecModelsWindowII} we show the mean model power spectra (purple curve), the mean true power spectra (black curve) and their corresponding residuals (red dotted curve). We show their corresponding mean fractional errors in purple in the third column. As discussed in Section \ref{sss:P1VConfig} the P1V visibilities are noisier than the simulated visibilities and contain instrument systematics not present in simulations. This manifests in the true power spectrum as increased amplitude for large $\tau$ modes, as well as the systematic feature at $\tau \pm 1.2 \mu$s. Referring to the model power spectra in the middle column of Figure \ref{fig:Sec7p4PspecModelsWindowII}, we can see that the inpainting techniques reproduce this systematic feature. Referring to the first row of the second column in Figure \ref{fig:Sec7p4PspecModelsWindowII} it appears that $P_{\rm model}$ for U-Paint have a similar amplitude as $P_{\rm true}$ for large $\tau$ modes. However referencing $P_{\rm model}$ for U-Paint with simulated data (upper left panel) shows that U-Paint models automatically produce this amplitude for large $\tau$. 

By referring to the mean fractional errors on the right column of Figure \ref{fig:Sec7p4PspecModelsWindowII} we can see that the mean fractional errors of each inpainting technique lie within the range $ 10^{-3} < \overline{\epsilon}^{\rm P}_{\rm frac} < 10$, where the largest fractional errors occur outside the wedge. The smallest fractional errors are again found for modes inside the wedge. In the wedge modes, the fractional errors are within a fraction of a percent of their true value. Quantitatively, we find that the inpainting techniques are within $1.24\%, 0.32\%, 1.24\%, 1.0\%, 0.25\%$ for U-Paint, CLEAN, LSSA, GPR and DPSS respectively. 

To generate the probability density function of the errors in the model power spectra, we construct two samples sets. One set using $\tau$ modes outside the wedge and another set comprised of $\tau$ modes inside the wedge. In each case we use model power spectra from LSTs with at least one flagged pixel. In the purple curves of Figure \ref{fig:PspecDistributionsWindowII} we show the errors in the model power spectra. In the top row of Figure \ref{fig:PspecDistributionsWindowII} (purple curve) we show the probability density functions of the residuals. We find that U-Paint produces the largest range of errors $\epsilon^{\rm P}_{\rm r}$, followed by DPSS, LSSA, GPR and CLEAN. In the second row in Figure \ref{fig:PspecDistributionsWindowII} (purple curve) we show the probability density functions of the fractional errors $\epsilon^{\rm P}_{\rm frac}$ constructed using only wedge modes. Comparing the mean of the fractional error distributions in the wedge modes of model power derived from P1V data to the mean fractional errors of model power spectra derived from simulated visibilities (blue curve), we find that there is an increase in $\overline{\epsilon}^{\rm P}_{\rm frac}$ using all inpainting techniques. The largest increase in mean fractional errors occurs in DPSS and CLEAN inpainting techniques. With the smallest increase in fractional errors using U-Paint. Conversely, if we construct $p(\epsilon^{\rm P}_{\rm frac})$ using only modes outside the wedge (bottom row in Figure \ref{fig:PspecDistributionsWindowII}) we find that the range of fractional errors decreases as compared to its equivalent distribution derived from simulated data (third row). This is due to there being lesser amounts of noise in the simulated data as compared to the P1V data, thereby exposing the spectral errors in the inpaint models. 

Using the fractional errors $\epsilon^{\rm P}_{\rm frac}$ we can establish a relationship between the mean fractional errors in the inpainted simulated visibilities and their corresponding power spectra. We proceed similarly as in Section \ref{ssec:RealisticSpecWinSec64}. In the bottom row of Figure \ref{fig:ScatterNflaggedWindowIIP1VSims} we show the relationship between the mean fractional errors in the visibilities  $\epsilon^{\rm V}_{\rm frac}$  and the mean fractional errors in the power spectrum $\epsilon^{\rm P}_{\rm frac}$. Comparing this to the top row of Figure \ref{fig:ScatterNflaggedWindowIIP1VSims} we demonstrate that the relationship between the mean fractional errors in the inpainted P1V data and their corresponding power spectra follow the same relationship as with the simulated data. This is important since it suggests that intuition and error characterisation drawn from the simulated visibilities in Section \ref{sssec:relationship_between_pspec_vis_errs} translates to P1V data. This result is perhaps not so surprising given that the fractional errors of the visibilities and power spectrum are described by the same probability profile for the P1V data visibilities and power spectra. Recall above the mean of the fractional error distributions for the power spectra of P1V data are larger (except for U-Paint) that the corresponding mean fractional errors using simulated data. Similarly in Section \ref{ssec:P1VResultsV} we found that there was an increase in $\epsilon^{\rm V}_{\rm frac}$ in the P1V data as compared to the simulated data. These increases essentially shift the center of the scatter plots in the bottom row of Figure \ref{fig:ScatterNflaggedWindowIIP1VSims} as compared to the top row (simulated data). In the future we would like to be able to predict the errors in P1V based on the error characterization in the simulated data. However,  although the relationship between these quantities remains the same between simulated and P1V, the centering of the distributions still need to be accounted for.

\section{Conclusion}
\label{sec:Conclusion}

As 21cm instruments continue to push towards a detection of the 21cm power spectrum, quantification of the errors introduced into the data analysis due to inpainting RFI corrupted data can no longer be ignored. In this paper we assessed the performance of existing inpainting techniques at restoring RFI flagged data. Our results are indicative of general trends, but not an exhaustive comparison. We also introduced our convolutional neural network U-Paint which we show to be capable of inpainting RFI corrupted data. Along with existing methods, we quantified the errors introduced in the data analysis pipeline due to RFI. We perform our error quantification analysis on simulated data as well as real data used in HERA's Phase 1 limits. We find that inpainting techniques which incorporate high wavenumbers in delay space in their modeling are best suited for inpainting over narrowband RFI. Our parameter choices for DPSS make DPSS best suited for inpainting over narrowband RFI while our parameter choices for LSSA make LSSA more flexible to wide RFI gaps and narrow RFI gaps. We find that with our fiducial parameters, DPSS and CLEAN provide the best performance for narrowband RFI while GPR provides the best performance for wideband RFI.  We also find that the error distributions in the phase component of the visibilites are log normally distributed. We find that these results hold in real data as well as simulated data. Further, we find that the standard deviation of the errors increases monotonically with increasing thermal noise of the simulated dataset. 

To characterize the errors that inpainting cause in the 21cm delay power spectrum, we propagate the inpainted visibilities to the 21cm power spectrum. We find that all inpainting techniques can reproduce the wedge modes of the delay power spectrum to within $10\%$ of the true values. Since the inpainting techniques are not capable of inpainting noise, the errors are greatest for the largest delay modes. Currently, systematics and noise prevent instruments from accurately measuring the amplitude of the power spectrum at the largest delay modes. However we show that in the future, as these effects are reduced, CLEAN and DPSS can most accurately reproduce the true power spectra at high delay. Quantitatively the errors reach the same order of magnitude as the noise. Conversely we find that U-Paint imparts artificial fine frequency structure into the visibilities which manifests as an increase in power at the highest delay modes. We also established a relationship between the mean fractional error in the model visibilities and the mean fractional errors in the model power spectrum. We find that this relationship is linear if we restrict the errors in the model power spectrum to only wedge modes. We also show that this is the case for both real and simulated data. Moving forward we have a better understanding of how the inpainting portion of the data analysis pipeline affect the 21cm power spectrum. This is another important step we must undertake on our continued path to make a detection of the 21cm power spectrum. 
\section*{Acknowledgements}
The authors are delighted to acknowledge helpful discussions with Lisa McBride, Saurabh Singh, Aaron Parsons, Bryna Hazelton, Paul LaPlante, Jonathan Pober, and Andrea Pallottini. N.K. gratefully acknowledges support from the MIT Pappalardo fellowship.



\section*{Data Availability}
The software code underlying this article will be shared on reasonable request to the corresponding author.

\bibliographystyle{mnras}
\bibliography{example}



\%appendix
\bsp	
\label{lastpage}
\end{document}